\documentclass[11pt]{article}

\usepackage[margin=1in]{geometry}  % set the margins to 1in on all sides
\usepackage{graphicx}              % to include figuresx
\usepackage{amsmath}               % great math stuff
\usepackage{amsfonts}              % for blackboard bold, etc
\usepackage{amsthm}                % better theorem environments
\usepackage{color}
\usepackage{tabulary}
\usepackage{caption}
\usepackage{amssymb}
\usepackage{mathrsfs}
\date{}

\begin{document}

%\title{Fuchsian Groups, Circularly Ordered Groups, and Dense Invariant Laminations on the Circle}
\title{Reduction for stochastic biochemical reaction networks with multiscale conservations \\ \vspace{2 mm} {\small This pre-print has been accepted for publication in SIAM Multiscale Modeling \& Simulation. The final copyedited version of this paper will be available at https://www.siam.org/journals/mms.php.
}}

\author{Jae Kyoung Kim 
\thanks{Department of Mathematical Sciences, Korea Advanced Institute of Science and Technology ({jaekkim@kaist.ac.kr})}
\and
Grzegorz A. Rempala 
\thanks{Division of Biostatistics and Mathematical Biosciences Institute, The Ohio State University ({rempala.3@osu.edu})}
\and
Hye-Won Kang
\thanks{Department of Mathematics and Statistics, University of Maryland, Baltimore County ({hwkang@umbc.edu})}
}

%\dedicatory{This paper is dedicated to the memory of William Thurston (1946--2012).}

\maketitle

\abstract{ Biochemical reaction networks frequently consist of species evolving on multiple timescales. Stochastic simulations of  such networks are often computationally challenging and therefore various methods have been developed to obtain sensible stochastic approximations on the timescale of interest. One of the rigorous and popular approaches is the multiscale approximation method for continuous time Markov processes. In this approach, by scaling species abundances and reaction rates, a family of processes parameterized by a scaling parameter is defined. The limiting process of this family is then used to approximate the original process. However, we find that such approximations become inaccurate when combinations of species with disparate abundances either constitute conservation laws or form virtual slow auxiliary species. To obtain  more accurate approximation in such cases, we propose here an appropriate  modification of the original  method. 
} 

%\begin{keywords}
%biochemical reaction networks, stochastic system, continuous-time Markov chain. multiscale approximation, singular perturbation theory, timescale separation
%\end{keywords}

%\textcolor{magenta}{$q_1$ 	$q_2$}\\
%\textcolor{cyan}{$l_1$	$l_2$}\\
\section{Introduction}

Biochemical reaction networks frequently evolve with disparate timescales. The simulations of the stochastic system describing such multi-scale biochemical reaction networks are extremely slow because the computation is predominantly spent on simulating fast reactions \cite{Cai2007, Gillespie2007, michelotti2013binning, coifman2008diffusion}. One approach to resolve this problem is using disparate timescales among species \cite{van1985elimination, pelevs2006reduction,Cotter:2011:CAM}. Fast species regulated by fast reactions will quickly equilibrate to a quasi-steady-state (QSS) while other species (slow species) will continue to evolve slowly on a different timescale (slow timescale). Thus, on the slow timescale, the fast species are assumed in QSS, which is determined by the evolution of slow species. By replacing the fast species with their QSS, we can derive the reduced stochastic system depending solely on the slow species. Such reduced system accurately approximates the slow timescale dynamics of the original full stochastic system with a much lower computational cost. 

However, in most systems with nonlinear reactions, deriving the exact QSS is difficult, and thus various approximations for QSS have been proposed  \cite{berglund2003geometric, Rao2003,vanden2003fast, Goutsias2005, Cao2005, haseltine2005origins, salis2005equation, lotstedt2006dimensional, Barik2008, Macnamara2008, michelotti2013binning, Cotter:2016:CAE}. Since typically the accuracy of such approximations has been investigated numerically due to the lack of analytical tools, their validity is difficult to fully establish. Indeed, recent studies have shown the potential inaccuracy of a popular approach based on a deterministically derived QSS (e.g. Michaelis-Menten function) \cite{Bundschuh2003a, Thomas2011, Thomas2012, Agarwal2012, kim2014validity, kim2015relationship}. These results indicate the need for justification of the QSS approximation using theoretical analysis \cite{liu2010analysis, givon2007strong, jahnke2011reduced}. 

One method allowing for a rigorous analysis is the multiscale approximation method, which was first introduced in \cite{Ball:2006:AAM} and further developed and systemized in \cite{Kang:2013:STM}. The method is based on the idea of scaling species abundances, reaction rate constants, and time with a common scaling parameter to define a family of processes indexed by the scaling parameter. The limit of the family is then used to approximate the original process on the timescale of interest. This multiscale approximation method has provided accurate approximate reduced models for various multiscale stochastic biochemical reaction networks, including the complex model of the heat shock response in E. coli \cite{Kang:2012:MAH, Kang:2013:STM, Kang:2014:CLT}. The multiscale approximation method allows for a rigorous analysis of the accuracy of the reduced model using theorems in stochastic analysis such as the law of large numbers and the martingale central limiting theorem \cite{Kang:2014:CLT}. Recently, this method was extended to study the chemical reaction-diffusion networks \cite{Pfaffelhuber:2015:SLS}. The scaling method developed for the multiscale approximation has also been used to derive various tools to study chemical reaction networks having multiscale nature, such as hybrid approximation and its simulation algorithms \cite{Ganguly:2015:EBS,Ganguly:2015:JAS,Hepp:2015:AHS}, parameter sensitivity analysis  \cite{Gupta:2013:UEP,Gupta:2014:SAS}, and the error analysis for stochastic numerical schemes \cite{Anderson:2012:WEA,Anderson:2012:MMC}. 

The current paper proposes the modified multiscale approximation method, which leads to accurate approximations for a broader class of multiscale stochastic biochemical reaction networks than the original method. Even though we concentrate, for the sake of simplicity, on two specific examples of networks, our proposed approach is seen to apply more broadly. The paper is organized as follows. In Section \ref{sec:2}, we briefly review the procedure of the original stochastic multiscale approximation using an example of the Michales-Menten enzyme kinetics. We also point out that the resulting reduced model does not accurately approximate the original model if the system has conservation laws involving species whose abundances are on disparate scales. To improve the accuracy, we propose a modification for the multiscale approximation method in Section \ref{sec:3}. In Section \ref{sec:4}, using an example of the genetic oscillatory system, we show that the stochastic multiscale approximation leads to an inaccurate approximation if the approximation uses a slow auxiliary variable, the combination of fast species whose abundances are on disparate scales. On the other hand, for such system, our modified multiscale approximation method leads to an accurate approximation.   In Section \ref{sec:col}, we summarize our results and discuss future work. The details of our analysis described in the main text are provided in the appendix.

%
%The introduction introduces the context and summarizes the
%manuscript. It is importantly to clearly state the contributions of
%this piece of work. The next two paragraphs are text filler,
%generated by the \texttt{lipsum} package.
%
%\lipsum[2-3]
%
%% The outline is not required, but we show an example here.
%The paper is organized as follows. Our main results are in
%Section \ref{sec:main}, our new algorithm is in Section \ref{sec:alg}, experimental
%results are in Section \ref{sec:experiments}, and the conclusions follow in
%Section \ref{sec:conclusions}.

\section{Stochastic multiscale approximation method}
\label{sec:2}
\begin{figure}[t]
  \centering
 \includegraphics[width=5in]{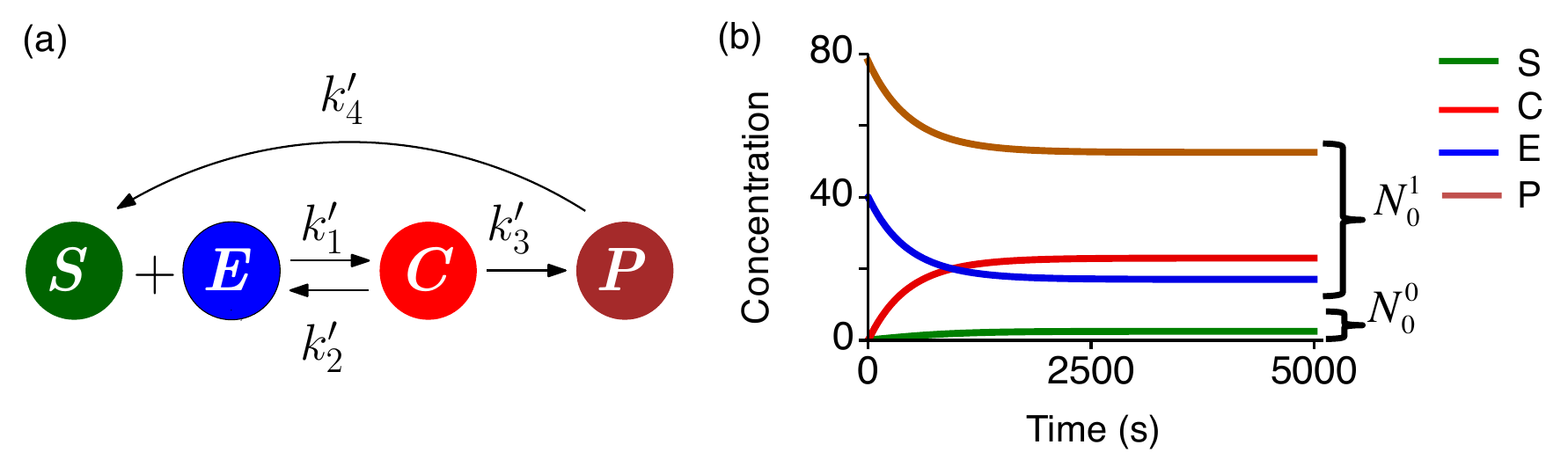}
\caption{Michaelis-Menten kinetics with a convertible product. (a) The diagram of the biochemical reaction network. (b) The simulations of ordinary differential equations, which are large volume limits of stochastic systems (\ref{orig}). When converting stochastic propensity functions to macroscopic reaction rates, volume $V=1/nM$ is assumed. Here, $S(0)=C(0)=0$, $E(0)=E_T (40nM) $, and $P(0)=S_T (80nM)$. For $N_0=10$, the scaling exponents for species abundance (\textit{i.e.} $\alpha_i$) are set to 0 for $S$ and 1 for others at the steady state.}
\label{det_simul}
\end{figure}

In this section, we review the multiscale approximation method \cite{Ball:2006:AAM,Kang:2012:MAH,Kang:2013:STM} and describe its limitations under conservation laws involving species with disparate molecular abundances. Consider a Michaelis-Menten enzyme kinetics with a product converting back to substrate \cite{Agarwal2012, kim2014validity}. This system consists of four reactions as described in \ref{det_simul}(a) and Table \ref{tab:prop}: a free enzyme ($E$) reversibly binds with a substrate ($S$) to form a complex ($C$) and then the complex irreversibly dissociates into a product ($P$) and a free enzyme. The product is assumed to be converted back to the substrate so that the substrate concentration is non-zero at the steady state. Propensity functions corresponding to these four reactions are derived based on the mass action kinetics by defining $X_i(t)$ be the abundance of the $i_{th}$ species at time $t$ (Table \ref{tab:prop}). 

\begin{table}[h!]
\begin{center}
\caption{Reactions and propensity functions of the Michaelis-Menten kinetics with a convertible product} \label{tab:prop}
  \begin{tabular}{| l | l | l |}  \hline
 \multicolumn{1}{|c|}{Reactions} & \multicolumn{1}{|c|}{Propensity functions}  \\ \hline
  $S + E\xrightarrow{\kappa'_1} C$ & $\lambda'_1 (X):=\kappa'_1 X_SX_E$\\
 $C \xrightarrow{\kappa'_2} S + E$ & $\lambda'_2 (X):=\kappa'_2X_C$  \\
 $C\xrightarrow{\kappa'_3} P + E$ & $\lambda'_3 (X):=\kappa'_3X_C$  \\
 $P \xrightarrow{\kappa'_4} S$ &$\lambda'_4 (X):=\kappa'_4X_P$  \\
   \hline  
\end{tabular}
\end{center}
\footnotesize{$\kappa_i'$ are stochastic reaction rate constants with units in the number of molecules rather than concentrations. $X_i(t)$ is the number of molecules of the $i_{th}$ species at time $t$.} 
\end{table}

Let $R^t_k (\cdot)$ be a counting process for the number of occurrences of the $k_{th}$ reaction up to time $t$ defined as 
\begin{equation}
R^t_k\left(\lambda'_k (X)\right):=Y_k \left(\int_{0}^{t} \lambda'_k (X(s))ds\right) \label{rk},
\end{equation}
where $Y_k$ are independent unit Poisson processes, and  $\lambda'_k(X)$ are the propensity functions of the $k_{th}$ reaction given in Table \ref{tab:prop}. With these counting processes, we can derive the system of stochastic equations describing the state of $X_i(t)$: 
\begin{equation}\label{orig}
\begin{split}
X_S (t) &= X_S (0) + R^t_2 (\lambda'_2 (X)) + R^t_4 (\lambda'_4(X)) -R^t_1 (\lambda'_1 (X)),\\
X_E (t) &= X_E(0)  + R^t_2 (\lambda'_2 (X)) + R^t_3 (\lambda'_3 (X)) - R^t_1 (\lambda'_1 (X)),\\ 
X_C (t) &= X_C (0) + R^t_1 (\lambda'_1 (X)) - R^t_2 (\lambda'_2 (X)) - R^t_3 (\lambda'_3 (X) ), \\
X_P (t) &= X_P (0) +  R^t_3 (\lambda'_3 (X)) - R^t_4 (\lambda'_4(X)).
\end{split}
\end{equation}
In this system, the total numbers of molecules of the substrate ($X_{S_T}$) and the enzyme ($X_{E_T}$) are conserved over time:
\begin{eqnarray} 
X_{S_T} &:=& X_S (t) +X_C (t) +X_P (t) = X_S (0) +X_C (0) +X_P (0), \label{xstcon}\\
X_{E_T} &:=& X_C (t) + X_{E} (t) = X_C (0) + X_{E} (0).
\end{eqnarray}
In the following subsections, we briefly describe how to derive the reduced system approximating the slow-scale dynamics of (\ref{orig}) with the multiscale approximation method \cite{Ball:2006:AAM,Kang:2012:MAH,Kang:2013:STM}.\\

\subsection{Deriving the normalized system}
The first step of the multiscale approximation method is scaling reaction rate constants, species abundances, and time via a common scaling parameter ($N_0$) to identify the timescale of each species. Here,  we choose the value of the scaling parameter as $N_0=10$ to transform the original reaction rate constants ($\kappa'_i$) to the normalized constants ($\kappa_i$) with $\kappa'_i=N^{\beta_i}_0 \kappa_i$. The scaling exponents ($\beta_i$) are chosen so that the normalized reaction rate constants ($\kappa_i$) are of order 1 as presented in Table \ref{tab:para}. 
\begin{table}[h!]
\begin{center}
\caption{Normalized reaction rate constants} \label{tab:para}
  \begin{tabular}{| l | l | l |l |}  
  \hline
   Name & Description  & Values \& Normalized rates ($\kappa_i$) \\ \hline
  $\kappa_1'$ & Binding rate constant for $E$ to $S$ & \hspace{0.1cm} $0.017/s=10^{-2}\times 1.7/s=:N_{0} ^{-2}\kappa_1$ \\
  $\kappa_2'$ & Unbinding rate constant for $C$  & \hspace{0.25cm} $0.03/s= 10^{-2}\times \hspace{0.25cm} 3/s   =: N_{0} ^{-2}\kappa_2$\\
 $\kappa_3'$ & Production rate constant for $P$  & $0.0016/s=10^{-3}\times 1.6/s =: N_{0} ^{-3}\kappa_3$ \\
 $\kappa_4'$ & Conversion rate constant for $P$ to $S$  & $0.0007/s=10^{-3}\times 0.7/s=: N_{0} ^{-3}\kappa_4$ \\
  \hline  
\end{tabular}
\end{center}
\footnotesize{The values of reaction rate constants are adopted from \cite{kim2014validity}} 
\end{table}

Similarly, the scaling exponents ($\alpha_i$) are chosen so that $X_i(t)/N^{\alpha_i}_0$ becomes of order 1. Since we are interested in the slow-scale dynamics of the system, we determine $\alpha_i$  based on the steady state values of the ordinary differential equations, which are the large volume limit (\textit{i.e.} thermodynamic limit) of the stochastic system  \cite{Kurtz:1972:RSD,gillespie2009JPC} (\ref{det_simul}(b)): 
\begin{eqnarray*}
\alpha_S = 0, \alpha_E = 1, \alpha_C= 1,  \alpha_P= 1.
\end{eqnarray*}

Using these scaling exponents, we define the normalized species abundance on the times of order $N_0^3$ as 
\begin{eqnarray} 
Z_i^{N_0}(t) &:=& \frac{X_i(t {N_0}^3)}{N_0^{\alpha_i}} \label{Z_norm1}
\end{eqnarray}
since we are interested in the dynamics at the timescale of order $N_0^3$ (\ref{det_simul}(b)). Then, we derive the counting processes in terms of the normalized rate constants ($\kappa_i$) and the normalized variables ($Z_i^{N_0}(t)$) on the timescale of order $N_0^3$. For instance, the counting process for the first reaction becomes  
\begin{eqnarray}
\begin{split}
Y_1 \left(\int_{0}^{N_0^3 t} \lambda'_1 (X(s))ds\right) &=Y_1 \left(\int_{0}^{N_0^3 t} \kappa_1' X_S (s) X_E (s) ds\right) \\ \label{gscale}
&=Y_1 \left(\int_{0}^{t} \left({N_0} ^{-2}\kappa_1\right) Z^{N_0}_S (u) \left(N_0 Z^{N_0}_E (u)\right) N_0^3 du\right)\\
&=:Y_1 \left(\int_{0}^{t} {N_0} ^{2} \lambda_1 (Z^{N_0}(u)) du\right),
\end{split}
\end{eqnarray}
where $Z^{N_0}$ is the vector whose $i_{th}$ component is $Z_i^{N_0}$. Here in the second equality, we apply the change of variable $s=N_0^3u$, and in the third equality, we define a normalized propensity function as $ \lambda_1 (Z^{N_0}) (u):=\kappa_1 Z^{N_0}_S (u) Z^{N_0}_E (u)$. In a similar way, we derive the counting processes for other reactions in terms of normalized propensity functions (see Table \ref{tab:npro}). Since $ \lambda_i (Z^{N_0})$ is of order 1, we can easily recognize the order of the counting processes in Table \ref{tab:npro}. The higher order indicates the faster counting process.  

\begin{table}[h!]
\begin{center}
\caption{Counting processes for the normalized system}\label{tab:npro}
  \begin{tabular}{| l | l | l |} 
  \hline
 Reaction & Counting processes  \\ \hline
 $S + E\xrightarrow{{N_0}^{-2}\kappa_1} C$ & $R^t_1 \left({N_0} ^{2}\lambda_1 (Z^{N_0})\right):=Y_1 \left(\int_{0}^{t} {N_0} ^{2} \kappa_1 Z^{N_0}_S (u) Z^{N_0}_E (u) du\right)$ \\
 $C \xrightarrow{{N_0}^{-2}\kappa_2} S + E$ & $R^t_2 \left({N_0} ^{2}\lambda_2 (Z^{N_0})\right):= Y_2 \left(\int_{0}^{t} {N_0} ^{2} \kappa_2 Z^{N_0}_C(u) du\right)$ \\
 $C\xrightarrow{{N_0}^{-3}\kappa_3} P + E$ & $R^t_3 \left({N_0}^{1}\lambda_3 (Z^{N_0})\right):= Y_3 \left(\int_{0}^{t} {N_0}^{1} \kappa_3 Z^{N_0}_C(u) du\right) $ \\
 $P \xrightarrow{{N_0}^{-3}\kappa_4} S$ & $R^t_4 \left({N_0} ^{1}\lambda_4 (Z^{N_0})\right):=Y_4 \left(\int_{0}^{t} {N_0}^{1}  \kappa_4 Z^{N_0}_P(u) du\right)$ \\
   \hline  
\end{tabular}
\end{center}
\footnotesize{  {Here, the scaling exponents, $\alpha_S = 0, \alpha_E = 1, \alpha_C= 1,  \alpha_P= 1$, are used to derive the normalized species abundance $Z_i^{N_0}$ as described in (\ref{Z_norm1}), and  the scaling exponents, $\beta_1 = -2, \beta_2 = -2, \beta_3= -3,  \beta_4= -3$ are used to derive normalized reaction rates as described in Table \ref{tab:para}. $\lambda_i(Z^{N_0})$ are normalized propensity functions for $i_{th}$ reactions, which are order of 1, and thus the orders of  reaction rates of $R_1, R_2, R_3$, and $R_4$ are 2, 2, 1, and 1, respectively.} } 
\end{table}

By substituting the counting processes in Table \ref{tab:npro} into the original stochastic system (\ref{orig}), we obtain the normalized stochastic system for $Z^{N_0}(t)$. In this normalized system, we replace now the fixed scaling parameter value $N_0$ with a varying parameter $N$ to derive a family of vector-valued processes $\{Z^{N} (t)\}$ depending on the parameter $N$:

\begin{equation}
\begin{split}
Z^{N}_S (t) &= Z^N_S (0) + R^t_2\left(N^2 \lambda_2(Z^{N})\right) + R^t_4 \left(N \lambda_4(Z^{N})\right) -R^t_1 \left(N^2 \lambda_1(Z^{N})\right),\\
Z^{N}_E (t) &= Z^N_E (0) + N^{-1} \left(R^t_2 \left(N^2 \lambda_2(Z^{N})\right) + R^t_3 \left(N \lambda_3(Z^{N})\right) -
 R^t_1 \left(N^2 \lambda_1(Z^{N})\right)\right),\\
Z^{N}_C (t) &= Z^N_C (0) + N^{-1} \left(R^t_1 \left(N^2 \lambda_1(Z^{N})\right) -R^t_2 \left(N^2 \lambda_2(Z^{N})\right) -R^t_3 \left(N \lambda_3(Z^{N})\right)\right) , \\
Z^{N}_P (t) &=Z^N_P (0) +  N^{-1} \left(R^t_3 \left(N \lambda_3(Z^{N})\right) - R^t_4 \left(N \lambda_4(Z^{N})\right)\right). \label{norms}
\end{split}
\end{equation}
The initial conditions for the family of precesses $\{Z^{N} (t)\}$ are defined so that $Z_i^{N}(0)\to Z_i^{N_0} (0)$ as $N\to\infty$:

\begin{equation}
\begin{split}
Z^{N}_S (0) &= Z^{N_0}_S (0) =X_S(0),\\ 
 Z^{N}_i(0) &= \frac{1}{N} \left\lfloor N Z^{N_0}_i(0)\right\rfloor =  \frac{1}{N} \left\lfloor\frac{N}{N_0}X_i(0)\right\rfloor, i=E, C, P. \label{intd}
\end{split}
\end{equation}
The floor function ($\lfloor~ \rfloor$) is used so that the initial conditions of unnormalized species $N^{\alpha_i}Z_i^N(0)$ have integer values (see \cite{Kang:2013:STM} for details). In the following, we will find the limit of this family of processes as $N \rightarrow \infty$ and use it to approximate the slow-scale dynamics of the stochastic system given in (\ref{orig}). Note that this approach is analogous to a singular perturbation approach based on Tikhonov's theorem \cite {tikhonov1952systems, Kepler2001, goeke2014constructive},  which reduces the multiscale \textit{deterministic} systems by setting a small scaling parameter as $0$ in the limit. \\

\subsection{Balance equations}

In the family of processes $\{Z^{N} (t)\}$ given in (\ref{norms}), the order of the maximum production rates for species $S$ is $N^2$ due to the term $R^t_2\left(N^2 \lambda_2(Z^{N})\right)$ since $\lambda_i (Z^N)$ is of order 1. The order of the maximum consumption rate  is also $N^2$ due to $R^t_1 \left(N^2 \lambda_1(Z^{N})\right)$. That is, both maximum production and consumption rates of species $S$ have the same scaling exponents as $2$. If the maximum exponent of the production rates is larger than that of the consumption rates, the normalized abundance of the species asymptotically goes to infinity as $N \to \infty$. In the opposite case, it asymptotically goes to zero in the limit. Thus, when the maximum exponents of production and consumption rates are equal, which is known as the ``balance equation", the limit of normalized species can be nondegenerate \cite{Kang:2012:MAH}. In case when there is a subset of species which do not satisfy the balance equations, their limit will be nondengenerate only for a certain time period, which gives the restriction on the choice of the timescale (see \cite{Kang:2012:MAH,Kang:2013:STM} for further details). In our example  in (\ref{norms}), all species and their linear combinations satisfy the balance equations. We also show that a nondegenerate limit of $\{Z^{N} (t)\}$ exists (see Appendix 2 for details).\\

\subsection{Deriving the average of fast variables and limiting model} 

For the species $P$ in (\ref{norms}), the maximum scaling exponent of the reaction rates and  the scaling exponent of species abundance (i.e. $\alpha_P$) are all $1$. This indicates that the number of molecules of $P$ and its change by reactions are of the same order on the current timescale, and therefore the current slow timescale is a natural timescale for $P$. In other words, $P$ is a slow-species in terms of the singular perturbation theory \cite{Kepler2001}. For other species,  $\alpha_i$ is less than the maximum scaling exponents of their reaction rates. Hence, the abundance of these species would fluctuate rapidly by reactions on the current slow timescale, indicating that they are fast species. Due to the rapid fluctuation, these fast species do not have a functional limit. Instead, they are averaged out in the limit as $N \to \infty$ \cite{Kurtz:1992:AMP,Ball:2006:AAM,Kang:2013:STM}. We now describe how to derive the average values of fast species in the limit. 

Using two conservation constraints of the systems (\ref{norms}):
\begin{eqnarray}
Z^{N}_{S_T}:&=&\frac{1}{N}Z^{N}_S(t)+Z^{N}_C(t)+Z^{N}_P(t)=\frac{1}{N}Z^{N}_S(0)+Z^{N}_C(0)+Z^{N}_P(0) \label{stcon},\\
Z^{N}_{E_T}:&=&Z^{N}_E(t)+Z^{N}_C(t)=Z^{N}_E(0)+Z^{N}_C(0)\label{etcon},
\end{eqnarray}
we can simplify (\ref{norms}) as 
\begin{eqnarray}
Z^N_S (t) &=& Z^N_S (0) + R^t_2\left(N^2 \kappa_2 Z_C^N \right) + R^t_4 \left(N \kappa_4Z_P^N\right) -R^t_1 \left(N^2 \kappa_1 Z_S^NZ_E^N\right), \label{zns_reduced}\\
Z^N_P (t) &= &Z^N_P (0) +  N^{-1} R^t_3 \left(N \kappa_3Z_C^N\right) - N^{-1} R^t_4 \left(N\kappa_4 Z_P^N\right)\label{znp_reduced}.
\end{eqnarray}
(\ref{zns_reduced})-(\ref{znp_reduced}) are closed since $Z_C^N(t)$ and $Z_E^N(t)$ are determined by $Z_S^N(t)$ and $Z_P^N(t)$ from the conservations in (\ref{stcon}-\ref{etcon}) as follows: 
\begin{eqnarray}
Z_C^N(t) &=&Z_{S_T}^N - \frac{1}{N}Z_S^N(t) - Z_P^N(t), \label{znc_reduced}\\ 
Z_E^N(t) &=& Z_{E_T}^N-Z_C^N(t) = Z_{E_T}^N-Z_{S_T}^N + \frac{1}{N}Z_S^N(t) + Z_P^N(t).  \label{zne_reduced}
\end{eqnarray}
Because the maximum order of the reaction rate ($N^2$) in (\ref{zns_reduced}) is greater than $N^{\alpha_S}=N^{0}$, species $S$ is rapidly fluctuating and thus its behavior in (\ref{znp_reduced}-\ref{znc_reduced}) is  averaged out as $N\to\infty$. To derive the averaged value, we use the law of large numbers for the Poisson process:
\begin{equation}
\lim_{N \to \infty} \sup_{x \leq x_0} \left|\frac{Y(N^{\alpha}x)}{N^{\alpha}}-x\right|=0 \label{plaw},
\end{equation}
where $\alpha>0, x_0>0$ and $Y$ is a unit Poisson process. From (\ref{plaw}), it follows that 
\begin{eqnarray*}
\frac{R^t_1 \left(N^2 \kappa_1 Z_S^NZ_E^N\right)}{N^2}&=&\frac{Y_1\left( \int_{0}^{t} N^{2}\kappa_1 Z^{N}_S(u) \left(Z_{E_T}^N-Z_{S_T}^N+\frac{1}{N}Z_S^N(u)+Z_P^N(u)\right)\,du\right)}{N^2} 
\end{eqnarray*}
has the same limit as the following integral:
\begin{eqnarray*}
\int_{0}^{t}\kappa_1 Z^{N}_S(u)\left(Z_{E_T}^N-Z_{S_T}^N+\frac{1}{N}Z_S^N(u)+Z_P^N(u)\right)\,du.
\end{eqnarray*}
Applying this result after dividing (\ref{zns_reduced}) by $N^2$, we get 
\begin{eqnarray*}
 \int_{0}^{t} \left(\kappa_2 Z_{C}^N(u)  - \kappa_1 Z_S^N(u)\left(Z_{E_T}^N-Z_{S_T}^N+\frac{1}{N}Z_S^N(u)+Z_P^N(u)\right) \right) \ du \nonumber
%\\&&=\int_{0}^{t} \left(  \kappa_2 \left(Z_{S_T}^N-\frac{1}{N}Z_S^N(u)-Z_P^N(u)\right)  - \kappa_1 Z_S^N(u)\left(Z_{E_T}^N-Z_{S_T}^N+\frac{1}{N}Z_S^N(u)+Z_P^N(u)\right) \right) \,du 
\rightarrow 0 
\end{eqnarray*}
as $N \to \infty$ since $Z^N_S (t)/N^2$ and $R^t_4 \left(N \kappa_4Z_P^N\right)/N^2$ go to zero. As  $Z^N_S (t)/N \to 0$ in the limit, we get
\begin{eqnarray}
&& \int_{0}^{t} \left(\kappa_2 Z_{C}^N (u)  - \kappa_1 Z_S^N{(u)}\left(Z_{E_T}^N-Z_{S_T}^N+Z_P^N(u)\right) \right) \ du \nonumber\\
&&=\int_{0}^{t} \left(  \kappa_2 \left(Z_{S_T}^N-Z_P^N(u)\right)  - \kappa_1 Z_S^N{(u)}\left(Z_{E_T}^N-Z_{S_T}^N+Z_P^N(u)\right) \right) \,du \rightarrow 0 \label{cave}
\end{eqnarray}
Setting the integrand of (\ref{cave}) to zero in the limit and defining $Z_P:=\lim_{N \to \infty} Z_P^N$, we can derive the  averaged value of the fast species ($\bar{Z}_S(t)$) in terms of the slow species ($Z_P(t)$) in the limit (see Appendix 1 for the detailed derivation): 
\begin{eqnarray}
\bar{Z}_S(t) &=& \frac{\kappa_2\left(Z_{S_T}-Z_P(t)\right)}{\kappa_1\left(Z_{E_T}-Z_{S_T}+Z_P(t)\right)}, \label{slim}
\end{eqnarray} 
where
\begin{eqnarray}
Z_{S_T} &=& \lim_{N \rightarrow \infty} Z^{N}_{S_T} = \frac{X_C(0)}{N_0}+ \frac{X_P(0)}{N_0}, \label{stconl} \\
Z_{E_T} &=& \lim_{N \rightarrow \infty} Z^{N}_{E_T} = \frac{X_E(0)}{N_0}+\frac{X_C(0)}{N_0} \label{etconl}.
\end{eqnarray}

Since $\bar{Z}_S(s)/N \to 0$ as $N \to \infty$, the  averaged value of another fast species ($C$) in the limit is also derived from (\ref{znc_reduced}) as
\begin{equation}
\bar{Z}_C(s) = Z_{S_T} - Z_P (s). \label{clim}
\end{equation}
Using this averaged value in the limit and the law of large numbers given in (\ref{plaw}), we get the limiting equation of (\ref{znp_reduced}): 
\begin{equation}
Z_P (t) =Z_P (0) +  \int_{0}^{t} \left( \kappa_3 \bar{Z}_C(s)- \kappa_4 Z_P(s) \right) \, ds. \label{plim}
\end{equation}
Note that this reduced system solely depends on $Z_P$ since $\bar{Z}_C(s)$ is determined by $Z_P(s)$ from (\ref{clim}). Following the original multiscale approximation method  \cite{Ball:2006:AAM,Kang:2013:STM}, we used $Z_P(t)$ of the limiting model to approximate $X_P(t)$ after unnormalizing the species abundance and rescaling back the time as
\begin{equation}
X_P (t) \approx N_0 Z_{P}(N_0^{-3}t) \label{papp}.
\end{equation}
The advantage of this approximation is that its error can be estimated using the law of large numbers and the martingale central limiting theorem \cite{Kurtz:1978:SAT,Kurtz:1981:APP,Ethier:1986:MPC,Kang:2014:CLT}. In our case, we get $X_P(t)=N_0 Z_P(N_0^{-3} t) + O(N_0^{1/2})$ since it has been known that $\frac{1}{N_0}X_P(N_0^3 t)-Z_P(t)=O\left(N_0^{-1/2}\right)$ \cite{Kang:2014:CLT}.  Note that  $X^{N}- Z^N = O(N^{-\beta})$ for some $\beta>0$ means that $ N^{\beta} \left(X^{N}(t)-Z^{N}(t)\right) \Rightarrow U(t)$ as $N \to \infty$ where $U(t)=O(1)$ (stochastically bounded). Here, $\Rightarrow$ indicates convergence in distribution (\textit{i.e.} weak convergence).

However, the approximation (\ref{papp}) obtained from the deterministic limiting model (\ref{plim}) cannot capture the fluctuation of $X_P (t)$. One natural way to resolve this issue is to replace the deterministic reaction terms in (\ref{plim}) by random jump processes with the corresponding propensity functions, which leads to the following stochastic process:
\begin{eqnarray}
\mathbb{Z}_P(t) &=& Z_P^{N_0}(0) + {N_0}^{-1} R_3^t\left( N_0 \kappa_3 \mathbb{Z}_C \right)
 - N_0^{-1} R_4^t\left( N_0\kappa_4 \mathbb{Z}_P \right), \label{approx1_ZPN}
\end{eqnarray}
where
\begin{eqnarray}
\mathbb{Z}_C(t) &=&   Z_{S_T} - \mathbb{Z}_P(t) \label{approx1_zcn}.
\end{eqnarray}
Note that this stochastic equation is the same as the original one for $Z_P^{N_0}$ in (\ref{znp_reduced}) except for $\mathbb{Z}_C(t)$, which now solely depends on the slow variable $ \mathbb{Z}_P(t)$ as $\bar{Z}_C(s)$ does in ($\ref{clim}$). Similarly to (\ref{papp}), we can use $\mathbb{Z}_P(t)$ in (\ref{approx1_ZPN}) to approximate $X_P(t)$, as $X_P(t) \approx N_0 \mathbb{Z}_P(N_0^{-3} t)$. 

\begin{figure}[t]
  \centering
\includegraphics[width=6.5in]{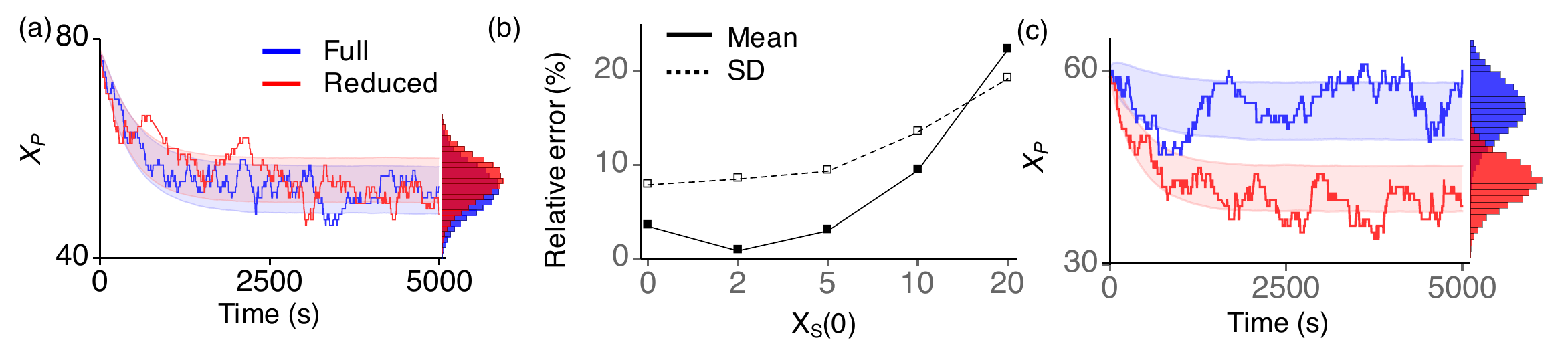}
\caption{The reduced model (\ref{approx1_ZPN}-\ref{approx1_zcn}) does not accurately approximate the original model  (\ref{orig}). (a)  The simulated trajectories of the original full model, $X_P(t)$, and the reduced model, $N_0 \mathbb{Z}_P(N_0^{-3} t)$. 
The colored ranges and histograms represent standard deviations of $X_P (t)$ and $N_0 \mathbb{Z}_P(N_0^{-3} t)$ from their mean and their distributions at the steady state, respectively. Here, the initial condition is the one used in \ref{det_simul}(b). In particular $X_S(0)=0$. (b) The relative differences of mean and standard deviation at the steady state (t=5000s) between the full model and the reduced model are numerically estimated for various values of $X_S (0)$. Here, $X_C (0)=0, X_E (0)=40, X_P(0)=80-X_S(0)$. (c)  The simulated trajectories of the original full model, $X_P(t)$, and the reduced model, $N_0 \mathbb{Z}_P(N_0^{-3} t)$ when $X_S(0)=20$. Due to the larger value of $X_S(0)$, the error becomes larger than (a).}
\label{fig:mmso}
\end{figure}

In Appendix 3, we show that 
\begin{eqnarray}
X_P(t) &\approx& N_0 \mathbb{Z}_P(N_0^{-3} t) +\mathbb{E}(N_0^{-3} t), \label{erro}\\
\mathbb{E}(t) &=&  \int_0^t \sqrt{ \kappa_3 \left| X_S(0) - \bar{Z}_S(s) -  \mathbb{E}(s) \right| 
+ \kappa_4 \left|\mathbb{E}(s)\right| } \,dW(s)\label{errt}\\
&& + \int_0^t \left\{  \kappa_3 \left( X_S(0) - \bar{Z}_S(s) -  \mathbb{E}(s) \right)  - \kappa_4 \mathbb{E}(s) \right\} \,ds, \nonumber
\end{eqnarray}
where $W$ is a standard Brownian motion. Importantly, $X_P(t) = N_0 \mathbb{Z}_P(N_0^{-3} t) + O(1)$ because $\mathbb{E}(t) = O(1)$, indicating that the new approximation with $N_0 \mathbb{Z}_P(N_0^{-3} t)$ is more accurate than the deterministic limit in (\ref{papp}). However, the new approximation with $N_0 \mathbb{Z}_P(N_0^{-3} t)$ still contains a considerable error as illustrated in Fig. \ref{fig:mmso}(a). In consistent with our error analysis in (\ref{errt}), the numerically estimated errors also increase as $\left| X_S(0) - \bar{Z}_S(s) \right|$ becomes larger considering the fact that $\bar{Z}_S(s) \approx 2$ (Fig. \ref{fig:mmso}(b) and (c)).

The dependence of errors on $\left| X_S(0) - \bar{Z}_S(s) \right|$ indicates that the error seen in Fig. \ref{fig:mmso} mainly stems from neglecting the species $S$ in the approximating process. 
Specifically, the initial condition of species $S$, $X_S(0)$, is ignored in the limiting total conserved quantity ($Z_{S_T}$) of (\ref{stconl}) due to the fact that the scaling exponent of $S$ ($\alpha_S$) is smaller than other scaling exponents in the conservation constraint (\ref{stcon}). For the same reason, $ \bar{Z}_S(s)$ is also neglected in the limit of the conservation constraint (\ref{clim}). Since $\bar{Z}_C(s)$ in (\ref{clim}) is used to derive (\ref{approx1_zcn}), $S$ is also neglected in the reduced model (\ref{approx1_ZPN}-\ref{approx1_zcn}). Therefore, as $X_S(0)$ takes a larger portion of $X_{S_T}$ in (\ref{xstcon}), ignoring $X_S(0)$ in deriving $Z_{S_T}$ causes a larger error as seen in Fig. \ref{fig:mmso}(b) and (c). 

Note that  we used one scaling exponent for species abundance of $S$ (\textit{i.e.} $\alpha_S=0$) for simplicity even when its order of magnitude of species abundance changes in time. In such case, $\alpha_S$ is supposed to be adjusted throughout time as suggested in the original multiscale approximation method  \cite{Kang:2012:MAH, Kang:2013:STM}. Specifically, when $X_S(0) = O(N_0)$ as in the case of Fig. \ref{fig:mmso}(c), it is suggested to use $\alpha_S=1$ for the initial transient period and $\alpha_S=0$ in the later time. However, with such multiple choices of $\alpha_S$ in time, the approximation process becomes complex since different reduced models will be derived in time and combining their numerical simulations is difficult.

\section{Modified multiscale stochastic approximation method}
\label{sec:3}
\begin{figure}[t]
  \centering
\includegraphics[width=6.5in]{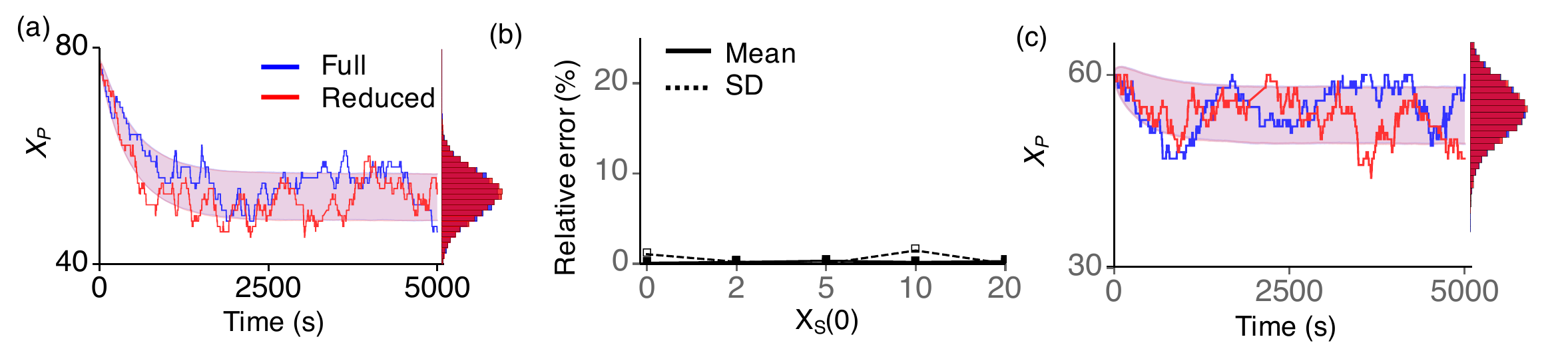}
\caption{The reduced model (\ref{plimf}) accurately approximates the original model (\ref{orig}). (a) The simulated trajectories of the original full model, $X_P(t)$, and the reduced model, $N_0\mathcal{Z}_P(N_0^{-3}t)$. The colored ranges and histograms represent the standard deviations of $X_P(t)$ and $N_0\mathcal{Z}_P(N_0^{-3}t)$ from their mean and their distributions at the steady state, respectively. The initial condition used in (a) is the same as those used in Fig. \ref{fig:mmso} (a). In particular, $X_S(0)=0$. (b) The relative differences of mean and standard deviation at the steady state (t=5000s) between the full model and the reduced model are numerically estimated for various $X_S (0)$. (c) The simulated trajectories of the original full model, $X_P(t)$, and the reduced model, $N_0\mathcal{Z}_P(N_0^{-3}t)$ when $X_S(0)=20$.  }
\label{fig:mmst}
\end{figure}

In order to correct the approximate errors seen in Fig. \ref{fig:mmso}, we introduce a modified conservation law of the normalized variables:

\begin{eqnarray}
\mathcal{Z}^{N}_{S_T}:&=&\frac{1}{N_0}Z^{N}_S(t)+Z^{N}_C(t)+Z^{N}_P(t)=\frac{1}{N_0}Z^{N}_S(0)+Z^{N}_C(0)+Z^{N}_P(0). \label{stcont}
\end{eqnarray}
Note that $\frac{1}{N}Z^{N}_S(t)$ in (\ref{stcon}) is replaced by $\frac{1}{N_0}Z^{N}_S(t)$ to prevent approximating $Z^{N}_S$ as 0 in the conservation law when $N \rightarrow \infty$. The limit of the newly derived total conserved quantity among the normalized species is 
\begin{eqnarray*}
\mathcal{Z}_{S_T} &:=& \lim_{N \to \infty} \mathcal{Z}^{N}_{S_T} =\frac{1}{N_0} \left( X_S(0) +X_C(0) +X_P(0) \right) =\frac{1}{N_0} X_{S_T}.
\end{eqnarray*}
In contrast to $Z_{S_T}$ in (\ref{stconl}), $\mathcal{Z}_{S_T}$ does not depend on the fraction of $X_S(0)$ in $X_S(0) +X_C(0) +X_P(0)$ as the total amount of the substrate, $X_{S_T}$, is fixed  \textemdash  $\mathcal{Z}_{S_T}$ is more natural conservation constant than $Z_{S_T}$. By substituting the new conservation constraint into (\ref{zns_reduced}-\ref{zne_reduced}), we define a new family of stochastic processes: 

\begin{eqnarray}
Z^N_S (t) &= & Z^N_S (0) + R^t_2 \left(N^2 \kappa_2 Z_C^N \right) + R^t_4 \left(N \kappa_4 Z_P^N \right) - R^t_1 \left(N^2\kappa_1 Z_S^N Z_E^N\right), \label{znst}\\
Z^N_P (t) &= & Z^N_P (0) +  N^{-1} R^t_3 \left(N\kappa_3 Z_C^N\right) - N^{-1} R^t_4 \left(N \kappa_4 Z_P^N \right),\label{znpt}\\
Z^N_C(t) &=& \mathcal{Z}^N_{S_T} - \frac{1}{N_0} Z^N_S(t) - Z^N_P(t), \label{stcontt}\\
Z^N_E (t) &=& Z^N_{E_T} - Z^N_C(t) = Z^N_{E_T} - \mathcal{Z}^N_{S_T} + \frac{1}{N_0} Z^N_S(t) + Z^N_P(t).\label{etcontt}
\end{eqnarray}
Though this new family of processes is different from the one in (\ref{zns_reduced}-\ref{zne_reduced}), we will use the same notation ($Z^N_i (t) $) for simplicity. Since (\ref{znst}-\ref{etcontt}) is equivalent to the original normalized system in (\ref{norms}) when $N=N_0$, the new family of processes includes the original system. Thus, the limiting model of (\ref{znst}-\ref{etcontt}) can be used to approximate the original system. To derive the limiting model, we divide (\ref{znst}) by $N^2$ and let $N \rightarrow \infty $ to get $\int_0^t \left(\kappa_2 Z^N_C(s)+\frac{1}{N}\kappa_4 Z_P^N(s) -\kappa_1 Z^N_S(s)Z^N_E(s)\right)\ ds \rightarrow 0$ in the same way as described in the previous section. As $\frac{1}{N}\kappa_4 Z_P^N(s) \to 0$, we get $\int_0^t \left(\kappa_2 Z^N_C(s)-\kappa_1 Z^N_S(s)Z^N_E(s)\right)\ ds \rightarrow 0$. Substituting (\ref{stcontt}-\ref{etcontt}) in the equation, we get
 \begin{eqnarray}
\int_0^t \left( \kappa_2 Z^N_C(s) -\kappa_1 N_0 \left(\mathcal{Z}^N_{S_T}  - Z^N_C(s)-Z^N_P(s)\right)\left(Z^N_{E_T}-Z^N_C(s)\right) \right)\, ds &\rightarrow& 0 \label{average}
\end{eqnarray}
as $N \rightarrow \infty $. %Using this, 
Setting the integrand to zero in the limit, we get the following approximation of the averaged value of fast species ($\overline{Z}_C$) with respect to the slow species $Z_P:=\lim_{N \to \infty} Z_P^N$:
 \begin{eqnarray}
 \bar{Z}_C(s) &\approx& \frac{Z_{E_T}+\mathcal{Z}_{S_T}-Z_P (s)+\frac{K_d}{N_0} }{2} \label{ZC_av} \\
&& -\frac{\sqrt{\left(Z_{E_T}+ \mathcal{Z}_{S_T} - Z_P(s)+\frac{K_d}{N_0}\right)^2-4Z_{E_T}\left(\mathcal{Z}_{S_T} - Z_P (s)\right) } }{2}, \nonumber
\end{eqnarray}
where $K_d=\frac{\kappa_2}{\kappa_1}$ (See Appendix 1 for detailed derivation). Using (\ref{ZC_av}) and the law of large numbers in (\ref{plaw}), and letting $N\to\infty$ in (\ref{znpt}), we get a limiting model for the slow species $P$:
 \begin{eqnarray}
Z_P (t) &\approx& Z_P (0) + \int_0^t \left( \kappa_3 \bar{Z}_C(s)- \kappa_4 Z_P(s) \right) \,ds. \label{ZP_hat} 
\end{eqnarray}
We convert this deterministic limiting model to the stochastic process as in the previous section: 
 \begin{equation}
\mathcal{Z}_P (t) = Z_P^{N_0} (0) + {N_0}^{-1} R^t_3\left({N_0} \kappa_3 \mathcal{Z}_C\right) - {N_0}^{-1} R^t_4\left({N_0} \kappa_4 \mathcal{Z}_P\right)\label{plimf},
\end{equation}
where 
\begin{eqnarray}
\mathcal{Z}_C(t) &=& \frac{Z_{E_T}+\mathcal{Z}_{S_T}-\mathcal{Z}_P(t)+\frac{K_d}{N_0} }{2} \label{climf} \\  
&&- \frac{\sqrt{ \left(Z_{E_T}+\mathcal{Z}_{S_T}-\mathcal{Z}_P(t)+\frac{K_d}{N_0}\right)^2 - 4 Z_{E_T} \left(\mathcal{Z}_{S_T}-\mathcal{Z}_P(t)\right) }}{2}.\nonumber
\end{eqnarray}
Note that in this new approximation, $\mathcal{Z}_C(t)$ is determined by $\mathcal{Z}_P (t)$ differently from the previous approximation in (\ref{approx1_ZPN}-\ref{approx1_zcn}). We again use $N_0\mathcal{Z}_P(N_0^{-3}t)$ to approximate $X_P(t)$ of the original model, which is accurate as seen in Fig. \ref{fig:mmst}(a). Furthermore, the new approximation is  accurate regardless of the initial condition of $S$ (Fig. \ref{fig:mmst}(b) and (c)) in contrast to the previous approximation (Fig. \ref{fig:mmso}). 

To investigate the accuracy of the new approximation, we perform the error analysis and obtain the following:
\begin{eqnarray}
X_P(t) &\approx& N_0\mathcal{Z}_P(N_0^{-3}t) + \mathcal{E}(N_0^{-3}t),\\
\mathcal{E}(t) &=&  \int_0^t \sqrt{\left(\kappa_3+\kappa_4\right) \left|\mathcal{E}(s)\right|} \,dW(s)
- \int_0^t \left(\kappa_3+\kappa_4\right) \mathcal{E}(s) \,ds, 
\end{eqnarray}
where $W$ is a standard Brownian motion  (see Appendix 4 for detailed analysis). In particular, since $\mathcal{E}(0)=0$ and the diffusion and drift terms are proportional to $\mathcal{E}(s)$, it follows that $\mathcal{E}(t)=0$ and  thus $X_P(t) =  N_0\mathcal{Z}_P(N_0^{-3}t) + o(1)$, which shows the accuracy of the newly reduced model in (\ref{plimf}-\ref{climf}). Note that $X^N=Z^N+o\left(N^{-\beta}\right)$ for some $\beta>0$ means that $N^{\beta}\left(X^N(t)-Z^N(t)\right) \Rightarrow 0$ as $N \to \infty$, where $\Rightarrow$ indicates convergence in distribution (\textit{i.e.} weak convergence).

\section{Multiscale approximation for a genetic oscillatory system}
\label{sec:4}
\begin{figure}[t]
  \centering
\includegraphics[width=5in]{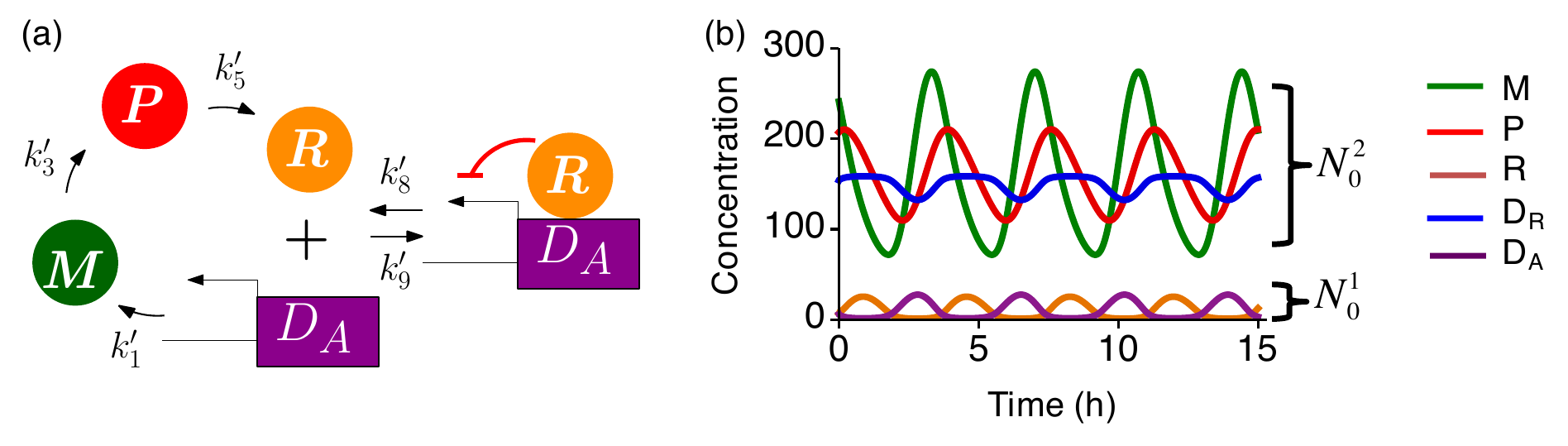}
\caption{Transcriptional negative feedback loop, (a) The diagram of the biochemical reaction network. (b) The simulations of ordinary differential equations, which is the large volume limit of stochastic system (\ref{osorig}).  When converting stochastic propensity functions to macroscopic reaction rates, volume $V=1nM^{-1}$ is assumed. Here, $M(0)=180nM, P(0)=210nM, R(0)=20nM, D_R(0)=160nM$, and $D_A(0)=0nM$. For $N_0=10$, the scaling exponents ($\alpha_i$) for species abundance become 1 for $R$ and $D_A$, and 2 for others. }
\label{fig:Osd}
\end{figure}

In the previous section, we propose a modified multiscale approximation method that leads to an accurate approximation for the stochastic system with a single steady state. In this section, we apply the same idea to the transcriptional negative feedback loop system, which generates oscillations (Fig. \ref{fig:Osd} (a)) \cite{Kim2012, kim2014validity, Kim2014, kim2016IET}. This system consists of 9 reactions as described in Table \ref{tab:oprop}: the transcription of mRNA ($M$) occurs proportional to active DNA ($D_A$) and then $M$ is translated into protein ($P$), which promotes the production of the repressor ($R$). The repressor reversibly binds with $D_A$ to form repressed DNA complex ($D_R$). Furthermore, $M$, $P$, and $R$ degrade. This model is described with the following set of stochastic equations: 
\begin{equation}\label{osorig}
\begin{split}
X_M(t) &= X_M(0) + R^t_1 (\lambda'_1(X)) - R^t_2 (\lambda'_2(X))\\
X_P(t) &=  X_P(0) + R^t_3 (\lambda'_3(X)) - R^t_4 (\lambda'_4(X))\\
X_R(t) &=  X_R(0) + R^t_5(\lambda'_5(X)) - R^t_6(\lambda'_6(X)) -R^t_8 (\lambda'_8(X)) +R^t_9 (\lambda'_9(X))\\
X_{D_R}(t) &= X_{D_R} (0) + R^t_8(\lambda'_8(X)) -R^t_9 (\lambda'_9(X))- R^t_7 (\lambda'_7(X)) \\
X_{D_A}(t) &= X_{D_A} (0) - R^t_8 (\lambda'_8(X)) +R^t_9 (\lambda'_9(X)) + R^t_7 (\lambda'_7(X)).
\end{split}
\end{equation}
Note that the total number of DNA ($X_{D_T}$) is conserved 
\begin{equation}
X_{D_T} := X_{D_A}(t) + X_{D_R}(t)= X_{D_A}(0) + X_{D_R}(0).
\end{equation}

\begin{table}[h!]
\begin{center}
\caption{Reactions and propensity functions} \label{tab:oprop}
  \begin{tabular}{| l | l | l |}  \hline
 Reactions & Original \& normalized propensity functions   \\ \hline
  $D_A \xrightarrow{\kappa'_1} D_A + M$ & $\lambda'_1 (X):=\kappa'_1X_{D_A}=N^{2}_0\kappa_1Z^{N_0}_{D_A}=:N^{2}_0\lambda_1 (Z^{N_0})$ \\
 $M \xrightarrow{\kappa'_2} \phi$ & $\lambda'_2 (X):=\kappa'_2X_M=N^{2}_0\kappa_2Z^{N_0}_{M}=:N^{2}_0\lambda_2 (Z^{N_0})$ \\
 $M \xrightarrow{\kappa'_3} M + P$ & $\lambda'_3 (X):=\kappa'_3X_M=N^{2}_0\kappa_3Z^{N_0}_{M}=:N^{2}_0\lambda_3 (Z^{N_0})$\\
 $P \xrightarrow{\kappa'_4}  \phi$ &$\lambda'_4 (X):=\kappa'_4X_P=N^{2}_0\kappa_4Z^{N_0}_{P}=:N^{2}_0\lambda_4 (Z^{N_0})$ \\
  $P \xrightarrow{\kappa'_5} P + R$ &$\lambda'_5 (X):=\kappa'_5X_P=N^{2}_0\kappa_5Z^{N_0}_{P}=:N^{2}_0\lambda_5 (Z^{N_0})$ \\
 $R \xrightarrow{\kappa'_6} \phi$ &$\lambda'_6(X):=\kappa'_6X_R=N^{1}_0\kappa_6Z^{N_0}_{R}=:N^{1}_0\lambda_6 (Z^{N_0})$  \\
 $D_R \xrightarrow{\kappa'_7} D_A$ &$\lambda'_7 (X):=\kappa'_7X_{D_R}=N^{2}_0\kappa_7Z^{N_0}_{D_R}=:N^{2}_0\lambda_7 (Z^{N_0})$ \\
 $D_A + R \xrightarrow{\kappa'_8} D_R$ &$\lambda'_8 (X):=\kappa'_8X_{D_A}X_R=N^{4}_0\kappa_8Z^{N_0}_{D_A}Z^{N_0}_{R}=:N^{4}_0\lambda_8 (Z^{N_0})$ \\
 $D_R \xrightarrow{\kappa'_9} D_A + R$ &$\lambda'_9 (X):=\kappa'_9X_{D_R}=N^{4}_0\kappa_9Z^{N_0}_{D_R}=:N^{4}_0\lambda_9 (Z^{N_0})$  \\
\hline 
\end{tabular} 
\end{center}
\footnotesize{{The $7_{th}$ reaction describes the degradation of $R$ bound to DNA. $\kappa'_1=15.1745$/hr,  $\kappa'_8=200/$hr, $\kappa'_9=50/$hr and other $\kappa'_i$ are $1/$hr, which are adopted from \cite{kim2014validity}. Thus, for $N_0=10$, $\kappa'_1=N^1_0\kappa_1$,  $\kappa'_i=N^2_0\kappa_i$ for $i=8$ and $9$, and $\kappa'_i=N^0_0\kappa_i$ for others so that $\kappa_i$  are of order 1. The scaling exponents ($\alpha_i$), 1  for $R$ and $D_A$, and 2 for others  are used  to derive normalized species $Z_i^{N_0}$, which are of order 1. Hence, the normalized propensity functions ($\lambda_i (Z^{N_0})$) are of order  1, and the orders of reaction rates can be easily derived from $\lambda'_i(Z^{N_0})$/$\lambda_i(Z^{N_0})$.}} 
\end{table}

To derive the normalized system of (\ref{osorig}), we scaled reaction rate constants with $N_0 =10$: $\kappa'_1=N^1_0\kappa_1$,  $\kappa'_i=N^2_0\kappa_i$ for $i=8$ and $9$, and $\kappa'_i=N^0_0\kappa_i$ for others as seen in Table \ref{tab:oprop}. According to the simulations of the deterministic system, which is the large volume limit of (\ref{osorig}), the scaling exponents of the molecular abundance ($\alpha_i$) can be chosen as $1$ for $X_{D_A}$ and $X_{R}$ and $2$ for other species (Fig. \ref{fig:Osd} (b)). Using $\alpha_i$, we define the normalized species abundance at the times of order $N_0^0$ as $Z_i^{N_0} (t) := X_i(t)/N_0^{\alpha_i}$.

\begin{figure}[t]
  \centering
\includegraphics[width=5in]{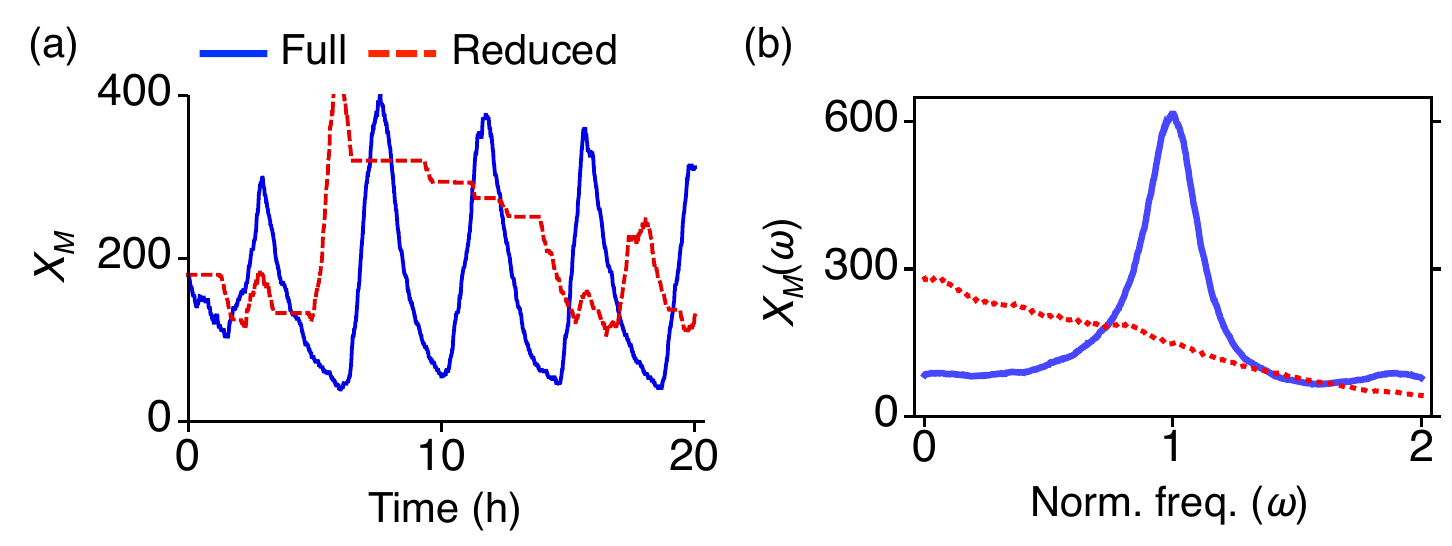}
\caption{ The reduced model (\ref{zmap}-\ref{zdaap}) does not accurately approximate the original full model (\ref{osorig}). (a) The simulated trajectories of the original full model, $X_M(t)$, and the reduced model, $N_0^2 \mathbb{Z}_M (t)$. The initial condition is the one used in Fig. \ref{fig:Osd} (b). (b) Fourier transforms of stochastic trajectories with $10^4$ cycles of the full and reduced model show a large difference.}
\label{fig:osso}
\end{figure}

Using the normalized species ($Z_i^{N_0} (t)$) and the normalized reaction rate constants ($\kappa_i$), we derive the normalized propensity functions ($\lambda_i (Z^{N_0})$), which are of order 1\,  as described in Table \ref{tab:oprop}. After replacing the original propensity functions in (\ref{osorig}) by the normalized ones, we replace $N_0$ with $N$ and obtain a family of vector-valued processes $\{ Z^N (t) \}$ satisfying
\begin{eqnarray*}
Z^{N}_M(t) &=& Z^{N}_M(0) + N^{-2}(R^t_1 (N^2\lambda_1(Z^N)) - R^t_2 (N^2\lambda_2(Z^N))),\\
Z^{N}_P(t) &=&  Z^{N}_P(0) + N^{-2}(R^t_3 (N^2\lambda_3(Z^N)) - R^t_4 (N^2\lambda_4(Z^N))),\\
Z^{N}_R(t) &=&  Z^{N}_R(0) + N^{-1}(R^t_5(N^2\lambda_5(Z^N)) - R^t_6(N\lambda_6(Z^N))  -R^t_8 (N^4\lambda_8(Z^N)) \\ &&+R^t_9 (N^4\lambda_9(Z^N))),\\
Z^{N}_{D_R}(t) &=& Z^{N}_{D_R} (0) + N^{-2}(R^t_8 (N^4\lambda_8(Z^N)) -R^t_9 (N^4\lambda_9(Z^N))- R^t_7 (N^2\lambda_7(Z^N))), \\
Z^{N}_{D_A}(t) &=& Z^{N}_{D_A} (0) +N^{-1}(-R^t_8 (N^4\lambda_8(Z^N)) +R^t_9 (N^4\lambda_9(Z^N)) + R^t_7 (N^2\lambda_7(Z^N))). 
\end{eqnarray*}
Initial conditions ($Z^{N}_{i}(0)$) are defined as done in the previous section (\ref{intd}). For all species, the exponents of the maximum production and consumption rates are the same (\textit{i.e.} balance equations are satisfied), justifying our choice of the timescale. Note that in the above system the normalized total DNA, $Z^{N}_{D_A} (t)/N + Z^{N}_{D_R}(t)$, is conserved. In the limit of this conserved relation, $Z^{N}_{D_A} (t)/N$ will be neglected, and thus all DNA is under repressed status in the limit. Thus, the reduced model with the original multiscale approximation method reaches the steady state rather than oscillates. This example again indicates that the limiting model derived using the original method does not accurately approximate the full model when the system has a conservation among species with disparate scales of molecular abundances. Thus, the modified conservation constraint as described in Section \ref{sec:3} is used as
\begin{eqnarray*}
Z^{N}_{D_T} &:=& Z^{N}_{D_A}(t)/N_0 + Z^{N}_{D_R}(t) = Z^{N}_{D_A}(0)/N_0 + Z^{N}_{D_R}(0),
\end{eqnarray*}
and the limit of $Z_{D_T}^N$ as $N\to\infty$ is defined as
\begin{eqnarray*}
Z_{D_T} &:=& \lim_{N\to\infty} Z_{D_T}^N =
 X_{D_A}(0)/N_0^2 + X_{D_R}(0)/N_0^2= X_{D_T}/N^2_0.
\end{eqnarray*}
Using this modified conservation constraint, we define a new family of stochastic processes, using the same notation ($Z^{N}_i(t)$) for simplicity:
\begin{eqnarray}
Z^{N}_M(t) &=& Z^{N}_M(0) + N^{-2}(R^t_1 (N^2\lambda_1(Z^N)) - R^t_2 (N^2\lambda_2(Z^N))), \label{nmdot}\\
Z^{N}_P(t) &=&  Z^{N}_P(0) + N^{-2}(R^t_3 (N^2\lambda_3(Z^N)) - R^t_4 (N^2\lambda_4(Z^N))),\label{nrcdot}\\
Z^{N}_R(t) &=&  Z^{N}_R(0) + N^{-1}(R^t_5(N^2\lambda_5(Z^N)) - R^t_6(N\lambda_6(Z^N)) -R^t_8 (N^4\lambda_8(Z^N))  \label{nrfdot} \\ 
 && +R^t_9 (N^4\lambda_9(Z^N))), \nonumber\\
Z^{N}_{D_R}(t) &=& Z^{N}_{D_R} (0) + N^{-2}(R^t_8 (N^4\lambda_8(Z^N))  -R^t_9 (N^4\lambda_9(Z^N))\label{ndrdot} \\ 
&&  - R^t_7 (N^2\lambda_7(Z^N))), \nonumber \\
Z^{N}_{D_A}(t) &=& N_0 (Z^{N}_{D_T}-  Z^{N}_{D_R}(t)).\label{ndadot}
\end{eqnarray}
Because the maximum scaling exponents of the reaction rates of species $R$ and $D_R$ are greater than the scaling exponents of molecular abundance ($\alpha_i$), $R$ and $D_R$ fluctuate rapidly and are  averaged out. To derive the average values of these fast variables, we divide (\ref{ndrdot}) by $N^2$ and use the law of large numbers for Poisson process in (\ref{plaw}) to  get
\begin{eqnarray}
&& \int_{0}^{t} \left(\kappa_8 Z^{N}_{D_A}(u) Z^{N}_{R}(u) -\kappa_9 Z^{N}_{D_R}(u)\right)du \nonumber\\
&&\qquad\qquad\qquad =\int_{0}^{t} \left(\kappa_8N_0 (Z^{N}_{D_T}-  Z^{N}_{D_R}(u))Z^{N}_{R}(u) -\kappa_9 Z^{N}_{D_R}(u)\right)du  \rightarrow 0 \label{rave}
\end{eqnarray}
as $N \rightarrow \infty$. Note that (\ref{rave}) consists of only the fast variables $Z_R$ and $Z_{D_R}$ and  thus, we cannot use (\ref{rave}) to derive the limiting average of the fast variables with respect to the slow variables. To circumvent  this problem, we introduce the auxiliary species $T=R+D_R$,  as suggested by the original multiscale approximation method \cite{Kang:2012:MAH,Kang:2013:STM}. Since the abundance of $T$ has  the same order as $D_R$, we get
\begin{equation}
Z^{N}_T (t):=(X_R(t)+X_{D_R}(t))/N^2 = N^{-1}Z^{N}_R(t) + Z^{N}_{D_R}(t) \label{znt},
\end{equation}
so that $Z^{N}_T (t)$ is of order 1. We now derive the equation for $Z^{N}_T (t)$ using (\ref{nrfdot})-(\ref{ndrdot}):
\begin{eqnarray}
Z^{N}_T(t) &=& Z^{N}_T(0) + N^{-2}(R^t_5(N^2\lambda_5(Z^N)) - R^t_{10}(N^2\lambda_{10}(Z^N))), \label{ntdot}\\
R_{10}(N^2\lambda_{10}(Z^N)) &:=& 
Y_6 \left(\int_{0}^{t} N\kappa_6Z^{N}_{R}(u) du \right)+Y_7\left(\int_0^t N^{2}\kappa_7Z^{N}_{D_R}(u) du \right) \nonumber\\
&\equiv&Y_{10} \left(\int_{0}^{t} \left(N\kappa_6Z^{N}_{R}(u)+N^{2}\kappa_7Z^{N}_{D_R}(u)\right) du \right)\nonumber\\
&=&Y_{10} \left(\int_{0}^{t} N^{2}\kappa_{10}Z^{N}_{T}(u) du\right)\nonumber.
\end{eqnarray}
Note that $\kappa_6=\kappa_7=1$ is used to define $\kappa_{10}:=\kappa_6=\kappa_7$, and thus  two reaction terms can be combined using the superposition principle of Poisson processes \cite{Durrett:2012:ESP}. The process for $Z^{N}_T(t)$ satisfies the balance equation, and $Z^{N}_T(t)$ is a slow variable because the maximum scaling exponent of the reaction rates and the scaling exponent for the species abundance are equal as $2$. We substitute (\ref{znt}) into (\ref{rave}) and get 
\begin{equation}
\int_{0}^{t} \left(\kappa_8N_0 (Z^{N}_{D_T}-  Z^{N}_{D_R}(u)) N(Z^{N}_T (u)-Z^{N}_{D_R}(u)) -\kappa_9 Z^{N}_{D_R}(u)\right)du \rightarrow 0 \label{nrave}
\end{equation}
as $N \rightarrow \infty$. Setting the integrand to zero in the limit, we derive the  averaged value of the fast species ($\bar{Z}_{D_R}$) in terms of the slow species in the limit ($Z_T(t):= \lim_{N \to \infty} Z^N_T(t)$): 
\begin{equation}
\bar{Z}_{D_R} (t)= Z_T(t)  \label{zdrlim}, 
\end{equation}
which is equivalent with the limit of (\ref{znt}).  (\ref{zdrlim}) { with} (\ref{ndadot}) yields the  averaged value of the fast species ($\bar{Z}_{D_A}$) 
\begin{equation}
\bar{Z}_{D_A} (t)= N_0(Z_{D_T} -Z_T(t)) \label{zdalim} .  
\end{equation}
Using $\bar{Z}_{D_A} (t)$ and the law of large number for the Poisson process, we get the limiting model for the slow species. Because the limiting model is deterministic, we convert it to the stochastic system similarly as we did in the previous section:
\begin{eqnarray}
\mathbb{Z}_M (t) &=& {Z}^{N_0}_M (0) + N_0^{-2} \left( R^t_1(N_0^2 \kappa_1 \bar{\mathbb{Z}}_{D_A})-R^t_2(N_0^2 \kappa_2 \mathbb{Z}_M) \right),  \label{zmap} \\
\mathbb{Z}_P (t) &=&{Z}^{N_0}_P (0) + N_0^{-2} \left( R^t_3(N_0^2 \kappa_3 \mathbb{Z}_{M})-R^t_4(N_0^2 \kappa_4 \mathbb{Z}_P)\right), \label{olimo} \\
\mathbb{Z}_T (t) &=&{Z}^{N_0}_T (0) + N_0^{-2} \left( R^t_5(N_0^2 \kappa_5 \mathbb{Z}_{P})-R^t_{10}(N_0^2 \kappa_{10} \mathbb{Z}_T)\right),\\
\bar{\mathbb{Z}}_{D_A}(t) &=& N_0 (Z_{D_T} -\mathbb{Z}_T(t))  \label{zdaap}.
\end{eqnarray}
Note that $\bar{\mathbb{Z}}_{D_A}(t)$ is derived from ($\ref{zdalim}$).  In Fig. \ref{fig:osso}, we used $\mathbb{Z}_M (t)$ to approximate $X_M (t)$ as $X_M(t) \approx N_0^2 \mathbb{Z}_M (t)$, but as seen from the plots, this approximation is inaccurate. In particular, the reduced model does not generate oscillations with a specific frequency in contrast to the full model (Fig. \ref{fig:osso}(b)) 

We wondered whether the inaccuracy  of the reduced model  (\ref{zmap}-\ref{zdaap}) stems frm the fact that we simply fixed scaling exponents ($\alpha_i=1$) for $R$ and $D_A$ throughout the oscillation as they change between $N_0^0$ and $N_0$  (Fig. \ref{fig:Osd}). That is, as $\alpha_i$ of $R$ and $D_A$  change throughout the oscillation, it might not be appropriate to fix the order of $\lambda'_8=\kappa'_8 X_{D_A}X_R$ as $N_0^4$ in Table \ref{tab:oprop}, which is used to derive  the equation for the average of fast species (\ref{nrave}). However, we find that although the orders of $X_{D_A}$ and { $X_{R}$} change, $\kappa'_8X_{D_A}X_R=O(N^{4}_0)$ throughout the oscillation. Thus our choice of fixed scaling exponents ($\alpha_i$) for $R$ and $D_A$ is not the reason for the inaccuracy of the average of fast species (\ref{zdrlim}) and thus the reduced model seen in Fig. \ref{fig:osso}.

Instead, we find that the inaccurate approximation of the  averaged value of the fast species in (\ref{zdaap}) is due to the fact that the slow auxiliary species ($T$) consists of fast species with disparate abundance scales and thus a fast species ($R$) with low scale of abundance is neglected in the limit. Specifically, $\bar{Z}_{D_R} (t) = Z_T(t)$ in (\ref{zdrlim}) is equivalent to approximating $N^{-1}Z^{N}_R(t)$ by $0$ in $Z^{N}_T (t)=Z^{N}_{D_R}(t) + N^{-1}Z^{N}_R(t)$ as $N \to \infty$. Since $\bar{Z}_{D_R} (t) = Z_T(t)$ is used to derive $\bar{Z}_{D_A} (t)$ in (\ref{zdalim}) and hence $\mathbb{Z}_{D_A}(t)$ in (\ref{zdaap}), $R$ is also neglected in the reduced system given in (\ref{zmap}-\ref{zdaap}), which leads to apparent errors seen in (Fig. \ref{fig:osso}). 
  
\begin{figure}[t]
  \centering
\includegraphics[width=5in]{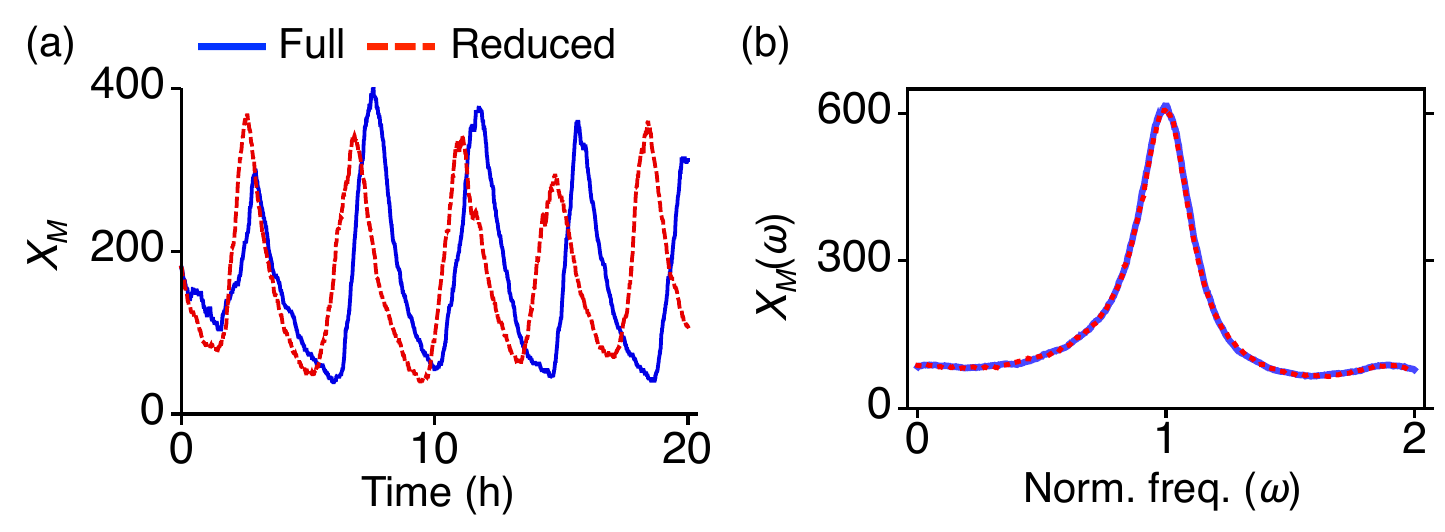}
\caption{  The reduced model  (\ref{zmapt}-\ref{zdaapt}) accurately approximate the original full model   (\ref{osorig}). (a) The simulated trajectories of the original full model,  $X_M(t)$, and  the reduced model, $N_0^2 \mathcal{Z}_M (t)$. The initial condition is the one used in Fig. \ref{fig:Osd} (b). (b) Fourier transforms of stochastic trajectories with $10^4$ cycles of the full and reduced model are consistent.}
\label{fig:osst}
\end{figure}

To resolve this problem, we adopt a similar idea to the one used in the previous section because a slow variable, $Z^N_T(t)$, is considered as a constant on fast timescale and thus (\ref{znt}) can be considered as a conservation law on fast timescale. We re-define $Z^N_T$ as
\begin{equation}
Z^{N}_T (t):=Z^{N}_{D_R}(t) + N_0^{-1}Z^{N}_R(t) \label{zntt},
\end{equation}
which prevents the elimination of $Z^{N}_R$ as $N \to \infty$. Though (\ref{zntt}) is different from (\ref{znt}), we keep using the notation $Z^{N}_T (t)$ for simplicity. With this new definition, we get the modified relation of (\ref{nrave}):
% \HWcom{We may need to derive (47) here or cite (44) and add explanation that we substitute $Z_R^N(u)$ by other variables using (54), if we decide to remove (47).}
\begin{equation}
\int_{0}^{t} \left(\kappa_8N_0 (Z^{N}_{D_T}-  Z^{N}_{D_R}(u)) N_0(Z^{N}_T (u)-Z^{N}_{D_R}(u)) -\kappa_9 Z^{N}_{D_R}(u)\right)du \rightarrow 0
 \label{nravet}
\end{equation}
as $N \rightarrow \infty$. 
Setting the integrand to zero in the limit, we get the approximation for the  averaged limiting value of $Z_{D_R}$ as %This leads the approximation for the  averaged value of $Z_{D_R}$ in the limit as
\begin{eqnarray*}
\bar{Z}_{D_R} (t)&\approx&\frac{Z_{D_T}+\frac{K_d}{N_0^2}+ Z_T(t) -\sqrt{(\frac{K_d}{N_0^2}-Z_{D_T} + Z_T(t) )^2 + 4Z_{D_T} \frac{K_d}{N_0^2}} }{2},
\end{eqnarray*}
where $K_d = \kappa_9/\kappa_8$. Using (\ref{ndadot}), we get
\begin{eqnarray*}
\bar{Z}_{D_A} (t)&\approx&N_0\frac{Z_{D_T}-\frac{K_d}{N_0^2}- Z_T(t) +\sqrt{(\frac{K_d}{N_0^2}- Z_{D_T} +Z_T(t) )^2 + 4Z_{D_T} \frac{K_d}{N_0^2}} }{2}. 
\end{eqnarray*}
By using the approximate averaged value ($\bar{Z}_{D_A}$) and the law of large numbers, we obtain the modified liming model for the slow species. Since the limiting model is deterministic, as before, we convert it to the following stochastic system. %the macroscopic reaction rates in this limiting model to random jump processes with the corresponding propensity functions to obtain the  following stochastic system:
\begin{eqnarray}
\mathcal{Z}_M (t) &=& Z_M^{N_0} (0) + N_0^{-2} \left( R^t_1(N_0^2 \kappa_1 \bar{\mathcal{Z}}_{D_A})-R^t_2(N_0^2 \kappa_2 \mathcal{Z}_M) \right),  \label{zmapt} \\
\mathcal{Z}_P (t) &=& Z_P^{N_0} (0) + N_0^{-2} \left( R^t_3(N_0^2 \kappa_3 \mathcal{Z}_{M})-R^t_4(N_0^2 \kappa_4 \mathcal{Z}_P) \right), \label{olimt} \\
\mathcal{Z}_T (t) &=& Z_T^{N_0} (0) + N_0^{-2} \left( R^t_5(N_0^2 \kappa_5 \mathcal{Z}_{P})-R^t_{10}(N_0^2 \kappa_{10} \mathcal{Z}_T) \right),\\
\bar{\mathcal{Z}}_{D_A}(t) &=& N_0\frac{Z_{D_T}-\frac{K_d}{N_0^2}- \mathcal{Z}_T(t) +\sqrt{(\frac{K_d}{N_0^2}- Z_{D_T} +\mathcal{Z}_T(t) )^2 -4Z_{D_T} \frac{K_d}{N_0^2}} }{2} 
 \label{zdaapt}.
\end{eqnarray}
Note that this newly derived reduced system is the same as the one in (\ref{zmap}-\ref{zdaap}) except for (\ref{zdaapt}). We used $\mathcal{Z}_M (t)$ to approximate $X_M (t)$ as $X_M(t) \approx N_0^2 \mathcal{Z}_M (t)$. As seen from the simulation (Fig. \ref{fig:osst}), the reduced model accurately approximates the original full model. 

We can often obtain slow auxiliary variables by combining fast variables because fast reactions could cancel each other as seen in (\ref{ntdot}). These newly derived slow variables play a critical role in deriving the reduced models in the multiscale stochastic approximation method \cite{Cao2005,E2005a,Kang:2013:STM}. If the slow normalized auxiliary species are derived as proposed in the original method (\ref{znt}), the constituent fast species of the auxiliary species are ignored in the limit if their scales of abundances ($\alpha_i$) are smaller than those of other constituent fast species.  This leads to considerable errors as seen in Fig. \ref{fig:osso}. On the other hand, our modification of the auxiliary variables given in (\ref{zntt}) prevents the fast species with small abundance being neglected in the limit and leads to  more accurate approximation as shown in Fig. \ref{fig:osst}. 

\section{Conclusion}
\label{sec:col}
Cells consist of diverse species whose abundances are on disparate scales. For instance, the concentrations of metabolites vary more than $10^6$ fold in \textit{E. coli}: the concentration of glutamate and adenosine are about $10^2\mu M$ and $10^{-4} \mu M$, respectively \cite{bennett2009absolute}. Thus, biochemical reaction networks often have conservation laws involving species with disparate abundance scales. Furthermore, the combination of fast species with disparate abundance scales can also form virtual slow auxiliary species that evolve slowly due to the cancelation of the fast reactions. In such cases, with the original multiscale approximation method, the constitute species with the low abundance are ignored in the conservation constraint or in the auxiliary species of limiting models as shown in (\ref{stconl}) or (\ref{zdrlim}). Therefore, the original multiscale approximation method \cite{Ball:2006:AAM,Kang:2013:STM} can lead to potential errors in the limiting models as seen in our examples (Fig. \ref{fig:mmso} and Fig. \ref{fig:osso}). To address this problem, we proposed here to replace the scaling parameter $N$ by the fixed value $N_0$ in the conservation constraints and auxiliary variables as we did in (\ref{stcont}) and (\ref{zntt}). Using these modified conservation constraints (or auxiliary variables), we redefined the family of the normalized stochastic processes in such a way that its limit provides accurate approximations for the full stochastic systems of the Michaelis-Menten kinetics (Fig. \ref{fig:mmst}) and the genetic oscillator (Fig. \ref{fig:osst}). This indicates that our modified method is applicable for a broader class of multiscale stochastic biochemical reaction networks than the original method. 

 When the abundances of species evolve across multiple scales over time, the original mutiscale approximation method may require time-dependent scaling exponent  $\alpha_i$ and thus lead to different reduced models over time \cite{Kang:2012:MAH}. In this case, the approximation process becomes complex as it requires combining different reduced models over time. On the other hand, our modified multiscale approximation method using the fixed $\alpha_i$ produces an accurate approximation in our example although some species abundances change over time (Fig. \ref{fig:mmst}(c)). It would be interesting future work whether our modified method is applicable to general systems where the scales of species abundances change over time.

Interestingly, the reduced models obtained using our methods coincide with those derived with the stochastic total quasi-steady state approximation (total QSSA) approach  \cite{Barik2008, Macnamara2008, kim2014validity, kim2015relationship}. Therefore, the error analysis used in our work can be also applied to validate the accuracy of the stochastic total QSSA, which has been up until now investigated mostly numerically. Another interesting application of our work can be extension of our method to approximate stochastic reaction-diffusion systems \cite{isaacson2006incorporating, erban2009stochastic, Kang:2012:NMC, Hu:2014:SAR, Pfaffelhuber:2015:SLS}.

\appendix

\section*{Appendix 1. Derivation of the spatial averages of fast species in Section \ref{sec:2} and Section \ref{sec:3}}
%Here, we provide the detailed derivation for average of fast species (\ref{slim}) in the Section 2.  %This is redundant with the title of this seciton
From the original full model described in (\ref{zns_reduced}-\ref{znp_reduced}), we derive a scaled generator of $z=(z_S,z_P)$ as
\begin{eqnarray}
A_Nf(z) &=& N^2 \kappa_1 \left(Z_{E_T}^N-Z_{S_T}^N + \frac{1}{N}z_S + z_P\right) z_S 
\left[ f\left(z-e_S\right) - f(z) \right] \label{gen}\\
&& +N^2 \kappa_2 \left(Z_{S_T}^N - \frac{1}{N}z_S - z_P \right) \left[ f\left(z+e_S\right) - f(z) \right] \nonumber\\ 
&& +N \kappa_3 \left(Z_{S_T}^N - \frac{1}{N}z_S - z_P \right) \left[ f\left(z+\frac{1}{N}e_P\right) - f(z) \right] \nonumber\\
&& +N \kappa_4 z_P \left[ f\left(z+e_S-\frac{1}{N}e_P\right) - f(z) \right]. \label{gef}
\end{eqnarray}
Define an occupational random measure of $Z_S^N$ as
\begin{eqnarray*}
\Gamma^N\left(D\times[0,t]\right) &=& \int_0^t 1_D \left(Z_S^N(s)\right) \,ds 
\end{eqnarray*}
in the space of measures $\nu$ on $\mathbb{Z}^+\times[0,\infty)$ such that $\nu(\mathbb{Z}^+\times[0,t])=t$ and $\mathbb{Z}^+$ is the set of natural number and zero. Denote the space of measures as
$\mathcal{L}\equiv \mathcal{L}(\mathbb{Z}^+)$.

Setting $f(z)=z_S$ in $(\ref{gen})$, we define a martingale
\begin{eqnarray}
M^N(t) &=& Z_S^N(t) -Z_S^N(0) \label{averaging} \\
&&- \int_{\mathbb{Z}^+ \times [0,t]} N^2\Bigg[  \kappa_2 \left(Z_{S_T}^N-\frac{1}{N}z_S-Z_P^N(s)\right) + \frac{1}{N}\kappa_4Z_P^N(s) \nonumber\\
&& \qquad\qquad - \kappa_1 z_S\left(Z_{E_T}^N-Z_{S_T}^N+\frac{1}{N}z_S+Z_P^N(s)\right) \Bigg] \,\Gamma^N\left(dz_S\times ds\right). \nonumber
\end{eqnarray}
\{$Z_P^N\}$ and $\{\Gamma^N\}$ are relatively compact in $D_{\mathbb{R}^+}([0,\infty))$ and $\mathcal{L}$, respectively, where $D_{\mathbb{R}^+}([0,\infty))$ is the space of cadlag functions with $\mathbb{R}^+$ values and $\mathcal{L}$ is the space of measures (see Appendix 2). Therefore, we can set $(Z_P,\Gamma)$ be a limit point of $\{(Z_P^N,\Gamma^N)\}$ in $D_{\mathbb{R}^+}([0,\infty))\times \mathcal{L}$.  Using Lemma 1.5 in \cite{Kurtz:1992:AMP}, 
\begin{eqnarray*}
&& \int_{\mathbb{Z}^+ \times [0,t]} \bigg[ \kappa_2 \left(Z_{S_T}^N-\frac{1}{N}z_S-Z_P^N(s)\right) + \frac{1}{N}\kappa_4Z_P^N(s) \\
&&\qquad\qquad - \kappa_1 z_S\left(Z_{E_T}^N-Z_{S_T}^N+\frac{1}{N}z_S+Z_P^N(s)\right) \bigg] \,\Gamma^N\left(dz_S\times ds\right)
\end{eqnarray*}
 converges in distribution to 
\begin{eqnarray}
&& \int_{\mathbb{Z}^+ \times [0,t]} \bigg[ \kappa_2 \left(Z_{S_T}-Z_P(s)\right) 
 - \kappa_1 z_S\left(Z_{E_T}-Z_{S_T}+Z_P(s)\right) \bigg] \,\Gamma\left(dz_S\times ds\right) \label{aveq}.
\end{eqnarray}

After dividing $(\ref{averaging})$ by $N^2$ and and letting $N$ go to infinity, the above term (\ref{aveq}) becomes zero for all $t>0$.
Using Lemma 1.4 in \cite{Kurtz:1992:AMP}, there exists $\mu_{(\cdot)}$ such that
$\Gamma(dz_S\times ds) = \mu_{Z_P(s)}(dz_S)\,ds$, and we get
\begin{eqnarray}
\int_{0}^{t}\int_{\mathbb{Z}^+} \left[  \kappa_2 \left(Z_{S_T}-Z_P(s)\right)  
- \kappa_1 z_S \left(Z_{E_T}-Z_{S_T}+Z_P(s)\right) \right] \,\mu_{Z_P(s)}(dz_S)\,ds &=& 0 \label{slim_eq_app}
\end{eqnarray}
with probability one.

Then, the average of fast species ($\bar{Z}_S$) is expressed in terms of the slow species ($Z_P$) as
\begin{eqnarray}
\bar{Z}_S(s) &\equiv& \int_{\mathbb{Z}^+} z_S \,\mu_{Z_P(s)}(dz_S) = \frac{\kappa_2\left(Z_{S_T}-Z_P(s)\right)}{\kappa_1\left(Z_{E_T}-Z_{S_T}+Z_P(s)\right)}, \label{slim_app}
\end{eqnarray}
which is given in the main text (\ref{slim}). Note that $\mu_{Z_P(s)}$ is a local-averaging distribution and the Poisson distribution with mean $\bar{Z}_S(s)$ because the limit of $A_Nf(z)/N^2$ in (\ref{gef}) is the infinitesimal generator of the Poisson process. For more details of conditions for averaging, please see Section 5 in \cite{Kang:2013:STM} and \cite{Ball:2006:AAM}.

Next, to derive the approximate averaged value of the fast species  (\ref{ZC_av}) of Section \ref{sec:3}, we substitute $\frac{1}{N}z_S$ to $\frac{1}{N_0}z_S$ and $Z_{S_T}^N$ to $\mathcal{Z}_{S_T}^N$ in (\ref{averaging}) and construct a new martingale corresponding to $Z_S^N$ in (\ref{znst})
\begin{eqnarray}
M^N(t) &=& Z_S^N(t) -Z_S^N(0) 
- \int_{\mathbb{Z}^+ \times [0,t]} N^2\Bigg[  \kappa_2 \left(\mathcal{Z}_{S_T}^N-\frac{1}{N_0}z_S-Z_P^N(s)\right)  \label{aver_hat_app}\\
&&\hspace{-0.5cm} + \frac{1}{N}\kappa_4Z_P^N(s)  
- \kappa_1 z_S\left(Z_{E_T}^N-\mathcal{Z}_{S_T}^N+\frac{1}{N_0}z_S+Z_P^N(s)\right) \Bigg] \,{\Gamma}^N\left(dz_S\times ds\right)\nonumber
\end{eqnarray}
where $\Gamma^N$ is an occupation measure of $Z_S^N$. 
$\left\{Z_P^N\right\}$ and $\left\{\Gamma^N\right\}$ are relatively compact, since $Z_P^N$ and $Z_S^N$ are bounded by $\mathcal{Z}_{S_T}^N\le \mathcal{Z}_{S_T}$ and $N_0\mathcal{Z}_{S_T}^N\le N_0\mathcal{Z}_{S_T}$ as seen in (\ref{stcont}), respectively.
%Dividing $(\ref{aver_hat_app})$ by $N^2$ and letting $N\to\infty$, $\{\hat{\Gamma}^N\}\Rightarrow \hat{\Gamma}$ with appropriate compactness conditions.
%Rewriting $\hat{\Gamma}(dz_S\times ds) = \hat{\mu}_{Z_P(s)}(dz_S)\,ds$ when $Z_P^N\toZ_P$ as $N\to\infty$,
Dividing $(\ref{aver_hat_app})$ by $N^2$ and taking a limit, we get
\begin{eqnarray*}
&& \int_0^t \int_{\mathbb{Z}^+} \Big[ \kappa_2 \left(\mathcal{Z}_{S_T} - \frac{1}{N_0}z_S - Z_P(s) \right) \\
&& \qquad\qquad\qquad   -\kappa_1 z_S \left(Z_{E_T}-\mathcal{Z}_{S_T} +\frac{1}{N_0}z_S + Z_P(s)\right)  \Big] \,{\mu}_{Z_P(s)}(dz_S)\,ds = 0
\end{eqnarray*}
as we derived (\ref{slim_eq_app}). Differentiating with respect to $t$ and replacing the time variable by $s$, the rewritten equation becomes
\begin{eqnarray*}
\int_{\mathbb{Z}^+} \left[ \frac{1}{N_0}z_S^2 + \left(Z_{E_T}-\mathcal{Z}_{S_T}+Z_P(s)+\frac{K_d}{N_0}\right) z_S 
- K_d\left(\mathcal{Z}_{S_T}-Z_P(s)\right) \right] \,{\mu}_{Z_P(s)}(dz_S) 
&=& 0 ,
\end{eqnarray*}
where $K_d=\frac{\kappa_2}{\kappa_1}$. 

We derive an approximate averaged value for $Z_S^N$ in the limit:

\begin{eqnarray*}
\int_{\mathbb{Z}^+} z_S\,{\mu}_{Z_P(s)}(dz_S) &\approx& -\frac{Z_{E_T}-\mathcal{Z}_{S_T}+Z_P(s)+\frac{K_d}{N_0}}{2/N_0} 
\label{moment_closure2_app}
\\
&&
+\frac{\sqrt{ \left(Z_{E_T}-\mathcal{Z}_{S_T}+Z_P(s)+\frac{K_d}{N_0}\right)^2 
+ \frac{4 K_d}{N_0}\left(\mathcal{Z}_{S_T}-Z_P(s)\right) }}{2/N_0} \mbox{} \nonumber
\end{eqnarray*}
by assuming $\int_{\mathbb{Z}^+} z_S^2 \,{\mu}_{Z_P(s)}(dz_S) \approx (\int_{\mathbb{Z}^+}z_S \,{\mu}_{Z_P(s)}(dz_S))^2$ in the limit. In the Appendix 4, we will show that this assumption does not cause any error up to the order of magnitude we are interest in.

\section*{Appendix 2. Relative compactness of \{$Z_P^N\}$ and $\{\Gamma^N\}$}

Here, we will show that \{$Z_P^N\}$ and $\{\Gamma^N\}$ in Appendix 1 are relatively compact in $D_{\mathbb{R}^+}([0,\infty))$ and $\mathcal{L}$, respectively, where $D_{\mathbb{R}^+}([0,\infty))$ is the space of cadlag functions with $\mathbb{R}^+$ values and $\mathcal{L}$ is the space of measures.  Since $Z_P^N(t)\le Z_{S_T}^N$ and $Z_{S_T}^N\to Z_{S_T}$ as $N\to\infty$, $Z_P^N(t)$ is bounded for all $t\in[0,\infty)$, and thus $\{Z_P^N(t)\}$ is relatively compact. We will show that for $t\in[0,\infty)$ and for fixed $\delta>0$, there exists $r$ such that
\begin{eqnarray*}
\sup_N P\left(\int_0^t 1_{[r,\infty)} \left(Z_S^N(s)\right)\,ds > \delta\right) &<& \delta.
\end{eqnarray*}
Since $\int_0^t 1_{[r,\infty)}\left(Z_S^N(s)\right)\,ds\le \int_0^t \frac{Z_S^N(s)}{r} \,ds$, we will show that we can set $P\left(\int_0^t \frac{Z_S^N(s)}{r}\,ds >\delta\right)$ small enough by choosing an appropriate value for $r$. We have
\begin{eqnarray*}
P\left(\int_0^t \frac{Z_S^N(s)}{r}\,ds>\delta\right) &\le& P\left(\inf_{t\in[0,\infty)} Z_E^N(t)\le \eta\right)
+ P\left(\int_0^t Z_S^N(s)Z_E^N(s)\,ds > r\delta\eta\right) \\
&\le& P\left(\inf_{t\in[0,\infty)} Z_E^N(t)\le \eta\right)
+ \frac{1}{r\delta\eta}E\left[\int_0^t Z_S^N(s)Z_E^N(s)\,ds \right].
\end{eqnarray*}
If $Z_E^N(0)\neq 0$ and $E\left[\int_0^t Z_S^N(s)Z_E^N(s)\,ds \right]<\infty$, we can set $\eta$ small enough and $r$ large enough so that both probabilities on the right-hand side become small. Then $Z_S^N(t)$ is stochastically bounded for $t\in[0,\infty)$, and by Lemma 1.1 in \cite{Kurtz:1992:AMP} $\{\Gamma^N\}$ is relatively compact. Now, we will show that $E\left[\int_0^t Z_S^N(s)Z_E^N(s)\,ds \right]<\infty$.
Taking the expectation on both sides of the equation for $Z_C^N(t)$ in (\ref{norms}) and rearranging terms, we have
\begin{eqnarray*}
E\left[\int_0^t \kappa_1 Z_S^N(s) Z_E^N(s)\,ds\right] &=& \frac{1}{N}E\left[Z_C^N(t)\right] -  \frac{1}{N}E\left[Z_C^N(0)\right] +E\left[\int_0^t\kappa_2 Z_C^N(s)\,ds\right] \\
&&+ \frac{1}{N}E\left[\int_0^t\kappa_3 Z_C^N(s)\,ds\right].
\end{eqnarray*}
The right-hand side is bounded since for all $t$, $Z_C^N(t)\le Z_{E_T}^N$ and this converges to $Z_{E_T}<\infty$ as $N\to\infty$.
%Assume that $Z_E^N(t)=Z_E^N(0)=0$ for all $t$. Then, $Z_C^N(t)=Z_C^N(0)>0$ for all $t$. This is only possible if 
%$\int_0^t\kappa_1Z_S^N(s)Z_E^N(s)\,ds=\int_0^t\kappa_2 Z_C^N(s)\,ds=\int_0^t\kappa_3Z_C^N(s)\,ds=0$ for all $t$. 
%This is contradictory to $Z_C^N(t)=Z_C^N(0)>0$ for all $t$.
%Therefore, we can find $\epsilon>0$ such that $Z_E^N(\epsilon)\neq 0$.
Note that we showed relative compactness of $\{\Gamma^N\}$ when $Z_E^N(0)\neq 0$. If $Z_E^N(0)=0$, we need additional assumption that $Z_S^N(t)$ is stochastically bounded for all $t\in[0,\infty)$.

\section*{Appendix 3. Error analysis for $\mathbb{Z}_P$ in Section \ref{sec:2}} 
To analyze the error of the process $\mathbb{Z}_P$ of (\ref{approx1_ZPN}) in approximating $Z_P^{N_0}$ of the full model in (\ref{znp_reduced}) with $N=N_0$, we use the technique developed in \cite{Anderson:2011:EAT}. To this end, we derive a family of process $\mathbb{Z}_P^N$ by replacing $N_0$ in (\ref{approx1_ZPN}) with the parameter $N$ as:
\begin{eqnarray}
\mathbb{Z}_P^N(t) &=& Z_P^N(0) + {N}^{-1} R_3^t\left( N \kappa_3 \mathbb{Z}_C^N \right)
 - N^{-1} R_4^t\left( N\kappa_4 \mathbb{Z}_P^N \right), \label{approx2_ZPN}
\end{eqnarray}
where
\begin{eqnarray}
\mathbb{Z}_C^{N}(t) &=&   \mathbb{Z}_{S_T}^N - \mathbb{Z}_P^N(t) \label{approx2_zcn}.
\end{eqnarray}
We define $\mathbb{Z}_{S_T}^N\equiv Z_C^N(0)+Z_P^N(0)$ so that $\mathbb{Z}_{S_T}^{N_0}=Z_{S_T}$. %and $\mathbb{Z}_C^{N_0}(t)=\mathbb{Z}_C(t)$. %We do not need this previous phrase. 
In this way, (\ref{approx2_ZPN}-\ref{approx2_zcn}) with $N=N_0$ become equivalent to the approximate model in (\ref{approx1_ZPN}-\ref{approx1_zcn}). Furthermore, 
 %$\mathbb{Z}_C^{N}(t) \equiv \mathbb{Z}_{S_T}^N - \mathbb{Z}_P^N(t)$ allows 
  $\mathbb{Z}_C^N(t)\to \bar{Z}_C(t)$ as $N\to\infty$ so that $\mathbb{Z}_P^N$ in (\ref{approx2_ZPN}) and $Z_P^N$ in (\ref{znp_reduced}) of the full model have the same limit $Z_P$ in $(\ref{plim})$. Since $Z_P^N(t) - \mathbb{Z}_P^{N}(t) \to 0$, we define an error between $Z_P^N$ and $\mathbb{Z}_P^N$ as
\begin{eqnarray}
\mathbb{E}^N(t) &\equiv& N\left(Z_P^N(t) - \mathbb{Z}_P^{N}(t)\right). \label{ent}
\end{eqnarray}
to get the asymptotic behavior of the error between $Z_P^N$ and $\mathbb{Z}_P^N$ of order $N^{-1}$.
To find an approximate value of $\mathbb{E}^{N_0}(t)$,
%\HWcom{`when $N_0$ is large'},
we derive a limiting behavior of $\mathbb{E}^N$ as $N\to\infty$. We rewrite the reaction terms for $Z_P^N$  in $(\ref{znp_reduced})$ as the following process, which has the same probability distribution with that in $(\ref{znp_reduced})$:

\begin{eqnarray}
Z_P^N(t) &=& Z_P^N(0) + \frac{1}{N} Y_{3,1}\left(\int_0^t N \kappa_3 Z_C^N(s) \wedge N\kappa_3 \mathbb{Z}_C^N(s) \,ds\right) \label{ZPN_2_app}\\
&&+ \frac{1}{N} Y_{3,2}\left(\int_0^t \left( N \kappa_3 Z_C^N(s) - N \kappa_3 Z_C^N(s) \wedge N\kappa_3 \mathbb{Z}_C^N(s) \right) \,ds\right) \nonumber\\
&&- \frac{1}{N} Y_{4,1}\left(\int_0^t N \kappa_4 Z_P^N(s) \wedge N \kappa_4 \mathbb{Z}_P^N(s) \,ds\right) \nonumber\\
&&- \frac{1}{N} Y_{4,2}\left(\int_0^t \left( N \kappa_4 Z_P^N(s)- N\kappa_4 Z_P^N(s) \wedge N \kappa_4 \mathbb{Z}_P^N(s)\right) \,ds\right), \nonumber
\end{eqnarray}
where $A\wedge B\equiv \min\left(A,B\right)$. Similarly, we rewrite the equation for $\mathbb{Z}_P^N$ in (\ref{approx2_ZPN}) as the following process:
\begin{eqnarray}
\mathbb{Z}_P^N(t) &=& Z_P^N(0) + \frac{1}{N} Y_{3,1}\left(\int_0^t N \kappa_3 Z_C^N(s) \wedge N\kappa_3 \mathbb{Z}_C^N(s) \,ds\right) \label{approx1_ZPN_2_app}\\
&&+ \frac{1}{N} Y_{3,3}\left(\int_0^t \left( N \kappa_3 \mathbb{Z}_C^N(s) - N \kappa_3 Z_C^N(s) \wedge N\kappa_3 \mathbb{Z}_C^N(s) \right) \,ds\right) \nonumber\\
&&- \frac{1}{N} Y_{4,1}\left(\int_0^t N \kappa_4 Z_P^N(s) \wedge N \kappa_4 \mathbb{Z}_P^N(s) \,ds\right) \nonumber\\
&&- \frac{1}{N} Y_{4,3}\left(\int_0^t \left(N \kappa_4 \mathbb{Z}_P^N(s)- N\kappa_4 Z_P^N(s) \wedge N \kappa_4 \mathbb{Z}_P^N(s)\right) \,ds\right). \nonumber
\end{eqnarray}
Subtracting $(\ref{approx1_ZPN_2_app})$ from $(\ref{ZPN_2_app})$,
\begin{eqnarray}
Z_P^N(t)-\mathbb{Z}_P^N(t) &=& 
\frac{1}{N} Y_{3,2}\left(\int_0^t \left( N \kappa_3 Z_C^N(s) - N \kappa_3 Z_C^N(s) \wedge N\kappa_3 \mathbb{Z}_C^N(s) \right) \,ds\right) \label{M1_diff1_app}\\
&&- \frac{1}{N} Y_{3,3}\left(\int_0^t \left( N \kappa_3 \mathbb{Z}_C^N(s) - N \kappa_3 Z_C^N(s) \wedge N\kappa_3 \mathbb{Z}_C^N(s) \right) \,ds\right) \nonumber\\
&&- \frac{1}{N} Y_{4,2}\left(\int_0^t \left( N \kappa_4 Z_P^N(s)- N\kappa_4 Z_P^N(s) \wedge N \kappa_4 \mathbb{Z}_P^N(s)\right) \,ds\right) \nonumber\\ 
&&+ \frac{1}{N} Y_{4,3}\left(\int_0^t \left(N \kappa_4 \mathbb{Z}_P^N(s)- N\kappa_4 Z_P^N(s) \wedge N \kappa_4 \mathbb{Z}_P^N(s)\right) \,ds\right). \nonumber
\end{eqnarray}

Taking the  reaction terms in $(\ref{M1_diff1_app})$ and subtracting  their propensity functions, we define the  following martingale
\begin{eqnarray*}
\mathbb{M}^{N}(t) 
&=&\frac{1}{N} \tilde{Y}_{3,2}\left(\int_0^t \left( N \kappa_3 Z_C^N(s) - N \kappa_3 Z_C^N(s) \wedge N\kappa_3 \mathbb{Z}_C^N(s) \right) \,ds\right)\\
&&- \frac{1}{N} \tilde{Y}_{3,3}\left(\int_0^t \left( N \kappa_3 \mathbb{Z}_C^N(s) - N \kappa_3 Z_C^N(s) \wedge N\kappa_3 \mathbb{Z}_C^N(s) \right) \,ds\right)\\
&&- \frac{1}{N} \tilde{Y}_{4,2}\left(\int_0^t \left( N \kappa_4 Z_P^N(s)- N\kappa_4 Z_P^N(s) \wedge N \kappa_4 \mathbb{Z}_P^N(s)\right) \,ds\right)\\ 
&&+ \frac{1}{N} \tilde{Y}_{4,3}\left(\int_0^t \left(N \kappa_4 \mathbb{Z}_P^N(s)- N\kappa_4 Z_P^N(s) \wedge N \kappa_4 \mathbb{Z}_P^N(s)\right) \,ds\right), 
\end{eqnarray*}
where $\tilde{Y}(u)=Y(u)-u$. A quadratic variation of the martingale is (\textit{cf}. {\cite{Kang:2014:CLT}) 
\begin{eqnarray*}
\left[\mathbb{M}^{N}\right]_t 
&=&\frac{1}{N^2} Y_{3,2}\left(\int_0^t \left( N \kappa_3 Z_C^N(s) - N \kappa_3 Z_C^N(s) \wedge N\kappa_3 \mathbb{Z}_C^N(s) \right) \,ds\right)\\
&&+ \frac{1}{N^2} Y_{3,3}\left(\int_0^t \left( N \kappa_3 \mathbb{Z}_C^N(s) - N \kappa_3 Z_C^N(s) \wedge N\kappa_3 \mathbb{Z}_C^N(s) \right) \,ds\right)\\
&&+ \frac{1}{N^2} Y_{4,2}\left(\int_0^t \left( N \kappa_4 Z_P^N(s)- N\kappa_4 Z_P^N(s) \wedge N \kappa_4 \mathbb{Z}_P^N(s)\right) \,ds\right)\\ 
&&+ \frac{1}{N^2} Y_{4,3}\left(\int_0^t \left(N \kappa_4 \mathbb{Z}_P^N(s)- N\kappa_4 Z_P^N(s) \wedge N \kappa_4 \mathbb{Z}_P^N(s)\right) \,ds\right). 
\end{eqnarray*}

Define a function for $Z_C^N$ in (\ref{znc_reduced}) and $\mathbb{Z}_C^N$ in (\ref{approx2_zcn}) as
\begin{eqnarray*}
F^N(z) &=& Z_{S_T}^N - \frac{1}{N}z_S - z_P \\
\bar{F}^N (z_P) &=& \mathbb{Z}_{S_T}^N -  z_P
\end{eqnarray*}
so that $F^N\left(Z^N(s)\right) = Z_C^N(s)$ and $\bar{F}^N \left( \mathbb{Z}_P^N(s)\right) = \mathbb{Z}_C^N(s)$.
As $N\to\infty$, $\left[\mathbb{M}^N\right]_t$ is asymptotic to
\begin{eqnarray*}
&&\frac{1}{N} \int_0^t \kappa_3 \left| Z_C^N(s) - \mathbb{Z}_C^N(s) \right| \,ds
+\frac{1}{N} \int_0^t \kappa_4 \left| Z_P^N(s)-\mathbb{Z}_P^N(s) \right| \,ds \\
&=&\frac{1}{N} \int_0^t \kappa_3 \bigg| \left[ F^N\left(Z^N(s)\right) - \bar{F}^N\left(Z_P^N(t)\right) \right]
+ \left[ \bar{F}^N\left(Z_P^N(t)\right) - \bar{F}^N\left( \mathbb{Z}_P^N(t)\right) \right] \bigg| \,ds \\
&& +\frac{1}{N} \int_0^t \kappa_4 \left| Z_P^N(s)-\mathbb{Z}_P^N(s) \right| \,ds, 
\end{eqnarray*}
where we use the fact that $\left(A-A\wedge B\right)+\left(B-A\wedge B\right) = \left|A-B\right|$. 
Then as $N\to\infty$, $\left[N\cdot \mathbb{M}^N\right]_t$ is asymptotic to
\begin{eqnarray}
&& \int_0^t \kappa_3 \bigg| N \left[ F^N\left(Z^N(s)\right) - \bar{F}^N\left(Z_P^N(t)\right) \right]
+ \frac{d \bar{F}^N \left(\mathbb{Z}_P^N(t)\right)}{d \mathbb{Z}_P^N(t)} \mathbb{E}^N(s)
\bigg| \,ds \label{MN1_asymptotic_app}.\\
&&+ \int_0^t \kappa_4 \left| \mathbb{E}^N(s) \right| \,ds \nonumber.
\end{eqnarray}

Subtracting and adding the propensity functions and using the fact that $\left(A-A\wedge B\right)-\left(B-A\wedge B\right) = \left(A-B\right)$, 
$(\ref{M1_diff1_app})$ can be rewritten as

\begin{eqnarray}
Z_P^N(t)-\mathbb{Z}_P^N(t) 
&=& 
\mathbb{M}^{N}(t) +\int_0^t \left[ \kappa_3 \left( Z_C^N(s) - \mathbb{Z}_C^N(s) \right)
- \kappa_4 \left( Z_P^N(s) - \mathbb{Z}_P^N(s) \right)\right] \,ds   \label{difference_2_app}\\
&=& 
\mathbb{M}^{N}(t) 
+\int_0^t \kappa_3 \left(  F^N\left(Z^N(s)\right) - \bar{F}^N\left(Z_P^N(s)\right) \right) \,ds  \nonumber\\
&&+ \int_0^t \kappa_3 \left( \bar{F}^N\left(Z_P^N(s)\right) - \bar{F}^N\left(\mathbb{Z}_P^N(s)\right) \right) \,ds \nonumber\\
&&- \int_0^t \kappa_4 \left( Z_P^N(s) - \mathbb{Z}_P^N(s) \right) \,ds. \nonumber
\end{eqnarray}
Multiplying $(\ref{difference_2_app})$ by $N$, we get
\begin{eqnarray}
\mathbb{E}^N(t) &\approx&  N\cdot \mathbb{M}^{N}(t) \label{E1N_app}\\
&&\hspace{-0.5cm} +\int_0^t \kappa_3 \left\{ N \left[  F^N\left(Z^N(s)\right) - \bar{F}^N\left(Z_P^N(s)\right) \right] 
+ \frac{d \bar{F}^N \left(\mathbb{Z}_P^N(s)\right)}{d \mathbb{Z}_P^N(s)} \mathbb{E}^N(s) \right\} \,ds \nonumber\\
&&\hspace{-0.5cm} - \int_0^t \kappa_4 \mathbb{E}^N(s) \,ds\nonumber
\end{eqnarray}
%

%\Jcom{Question!! Do we have to assume "$\mathbb{E}^N\Rightarrow \mathbb{E}$"? Isn't it guaranteed by the relative compactness proof?} \HWcom{I only assumed $\mathbb{E}^N\Rightarrow \mathbb{E}$ to explain the convergence of each term in (76)-(77). I am not sure how the relative compactness proof can be used here. For the final convergence equation for $\mathbb{E}$, we may not need this assumption.}
Assuming that $\mathbb{E}^N\Rightarrow \mathbb{E}$ as $N\to\infty$, where  $\Rightarrow$ implies convergence in distribution (or weak convergence), we get 
\begin{eqnarray}
&& N \left[  F^N\left(Z^N(s)\right) - \bar{F}^N\left(Z_P^N(s)\right) \right] \label{F_conv_app}\\
&=& N \left[ Z_{S_T}^N - \frac{1}{N}Z_S^N(s) - Z_P^N(s) - \mathbb{Z}_{S_T}^N + Z_P^N(s) \right] \nonumber\\
&=& N \left[  X_S(0)/N - Z_S^N(s)/N \right] \nonumber \\
&\longrightarrow& X_S(0) - \bar{Z}_S(s) \nonumber
\end{eqnarray}
and
\begin{eqnarray}
\frac{d \bar{F}^N \left(\mathbb{Z}_P^N(s)\right)}{d \mathbb{Z}_P^N(s)} \mathbb{E}^N(s)
&\longrightarrow&  -\mathbb{E}(s). \label{dF_conv_app}
\end{eqnarray}
Substituting $(\ref{F_conv_app})$ and $(\ref{dF_conv_app})$ to $(\ref{MN1_asymptotic_app})$ and applying the martingale central limit theorem, $N\cdot  \mathbb{M}^{N}\Rightarrow \mathbb{M}$ as $N\to\infty$, where $\mathbb{M}$ is a Gaussian process with its quadratic variation
\begin{eqnarray*}
\left[\mathbb{M}\right]_t &=& \int_0^t \left\{ \kappa_3 \left| X_S(0) - \bar{Z}_S(s) -  \mathbb{E}(s) \right| 
+ \kappa_4 \left|\mathbb{E}(s)\right| \right\} \,ds.
\end{eqnarray*}
Therefore, as $N\to\infty$, $(\ref{E1N_app})$ converges in distribution to
\begin{eqnarray*}
\mathbb{E}(t) &=&  \int_0^t \sqrt{ \kappa_3 \left| X_S(0) - \bar{Z}_S(s) -  \mathbb{E}(s) \right| 
+ \kappa_4 \left|\mathbb{E}(s)\right| } \,dW(s) \\
&& + \int_0^t \left\{  \kappa_3 \left( X_S(0) - \bar{Z}_S(s) -  \mathbb{E}(s) \right)  - \kappa_4 \mathbb{E}(s) \right\} \,ds, 
\end{eqnarray*} 
where $W$ is a standard Brownian motion and thus $\mathbb{E}(t)=O(1)$. Approximating $\mathbb{E}^{N_0}(t)\approx \mathbb{E}(t)$ as suggested in  \cite{Kang:2014:CLT} and using (\ref{ent}), we obtain
\begin{eqnarray*}
X_P(t) &\approx& N_0\mathbb{Z}_P(N_0^{-3}t) + \mathbb{E}(N_0^{-3}t),
\end{eqnarray*}
which indicates that $X_P(t)= N_0\mathbb{Z}_P(N_0^{-3}t) + O(1)$.

\section*{Appendix 4. Error analysis for $\mathcal{Z}_P$ in Section \ref{sec:3}} \label{error_anal2}
We again use the technique developed in \cite{Anderson:2011:EAT} to derive the error between $\mathcal{Z}_P$ of the approximate model (\ref{plimf}) and $Z_P^{N_0}$ of the full model (\ref{znp_reduced}) with $N=N_0$. To this end, we derive a family of the processes $\mathcal{Z}_P^N$ by replacing $N_0$ of $\mathcal{Z}_P$ in (\ref{plimf}) by a parameter $N$ as:

\begin{eqnarray}
\mathcal{Z}_P^{N}(t) &=& Z_P^N(0) + N^{-1} R_3^t\left( N\kappa_3 \mathcal{Z}_C^N \right) 
- N^{-1} R_4^t\left( N\kappa_4 \mathcal{Z}_P^{N} \right) \label{apnap},
\end{eqnarray}
where
\begin{eqnarray*}
 \mathcal{Z}_C^N(s) &=& \frac{Z_{E_T}^N+Z_{S_T}^N-\mathcal{Z}_P^N(s)+\frac{K_d}{N}}{2} \\
&& \qquad -\frac{\sqrt{ \left(Z_{E_T}^N+Z_{S_T}^N-\mathcal{Z}_P^N(s)+\frac{K_d}{N}\right)^2 
- 4Z_{E_T}^N \left(Z_{S_T}^N-\mathcal{Z}_P^N(s)\right) }}{2} .
\end{eqnarray*}
Note that  $\mathcal{Z}_C^{N_0}(t)=\mathcal{Z}_C(t)$ since $Z_{S_T}^{N_0}=\mathcal{Z}_{S_T}$. Then, $\mathcal{Z}_P^{N} (t)$ of (\ref{apnap}) when $N=N_0$ becomes equivalent to $\mathcal{Z}_P$ of (\ref{plimf}). That is, the family of process ($\mathcal{Z}_P^N$) includes the approximate process $\mathcal{Z}_P$ of (\ref{plimf}). Since $\mathcal{Z}_C^N(t)\to \bar{Z}_C(t)$ in (\ref{clim}) as $N\to\infty$, $\mathcal{Z}_P^N(t)$ and $Z_P^N(t)$ of the full model in (\ref{znp_reduced}) converge to the same limit $Z_P(t)$ in (\ref{plim}) as $N\to\infty$. Since $Z_P^N-\mathcal{Z}_P^N \to 0$ as $N\to\infty$, we define an error as
\begin{eqnarray*}
\mathcal{E}^N(t) &\equiv& N\left(Z_P^N(t) - \mathcal{Z}_P^{N}(t)\right)
\end{eqnarray*}
to get the asymptotic behavior of the error of order $\frac{1}{N}$ in $Z_P^N(t)-\mathcal{Z}_P^N(t)$.

To find an approximate of $\mathcal{E}^{N_0}(t)$, we investigate an asymptotic behaviour of $\mathcal{E}^N$ as $N\to\infty$. As we derived $(\ref{M1_diff1_app})$, we derive the following equation after replacing $\mathbb{Z}_P^N$ and $\mathbb{Z}_C^N$ by $\mathcal{Z}_P^N$ and $\mathcal{Z}_C^N$ 
in $(\ref{M1_diff1_app})$.
\begin{eqnarray}
Z_P^N(t)-\mathcal{Z}_P^N(t) &=& 
\frac{1}{N} Y_{3,2}\left(\int_0^t \left( N \kappa_3 Z_C^N(s) - N \kappa_3 Z_C^N(s) \wedge N\kappa_3 \mathcal{Z}_C^N(s) \right) \,ds\right) \label{M2_diff1_app}\\
&&- \frac{1}{N} Y_{3,3}\left(\int_0^t \left( N \kappa_3 \mathcal{Z}_C^N(s) - N \kappa_3 Z_C^N(s) \wedge N\kappa_3 \mathcal{Z}_C^N(s) \right) \,ds\right) \nonumber\\
&&- \frac{1}{N} Y_{4,2}\left(\int_0^t \left( N \kappa_4 Z_P^N(s)- N\kappa_4 Z_P^N(s) \wedge N \kappa_4 \mathcal{Z}_P^N(s)\right) \,ds\right) \nonumber\\ 
&&+\frac{1}{N} Y_{4,3}\left(\int_0^t \left(N \kappa_4 \mathcal{Z}_P^N(s)- N\kappa_4 Z_P^N(s) \wedge N \kappa_4 \mathcal{Z}_P^N(s)\right) \,ds\right). \nonumber
\end{eqnarray}
Using reaction terms in $(\ref{M2_diff1_app})$ and subtracting them by their propensity functions, define a martingale as
\begin{eqnarray*}
\mathcal{M}^{N}(t) 
&\equiv& \frac{1}{N} \tilde{Y}_{3,2}\left(\int_0^t \left( N \kappa_3 Z_C^N(s) - N \kappa_3 Z_C^N(s) \wedge N\kappa_3 \mathcal{Z}_C^N(s) \right) \,ds\right)\\
&&- \frac{1}{N} \tilde{Y}_{3,3}\left(\int_0^t \left( N \kappa_3 \mathcal{Z}_C^N(s) - N \kappa_3 Z_C^N(s) \wedge N\kappa_3 \mathcal{Z}_C^N(s) \right) \,ds\right)\\
&&- \frac{1}{N} \tilde{Y}_{4,2}\left(\int_0^t \left( N \kappa_4 Z_P^N(s)- N\kappa_4 Z_P^N(s) \wedge N \kappa_4 \mathcal{Z}_P^N(s)\right) \,ds\right)\\ 
&&+ \frac{1}{N} \tilde{Y}_{4,3}\left(\int_0^t \left(N \kappa_4 \mathcal{Z}_P^N(s)- N\kappa_4 Z_P^N(s) \wedge N \kappa_4 \mathcal{Z}_P^N(s)\right) \,ds\right),
\end{eqnarray*}
where $ \tilde{Y}(u)=Y(u)-u$. Define 
\begin{eqnarray*}
\tilde{F}^N\left(z_P\right)  &\equiv& \frac{Z_{E_T}^N+Z_{S_T}^N-z_P+\frac{K_d}{N}}{2} \nonumber\\
&&  -\frac{\sqrt{ \left(Z_{E_T}^N+Z_{S_T}^N-z_P+\frac{K_d}{N}\right)^2 
- 4Z_{E_T}^N \left(Z_{S_T}^N-z_P\right) }}{2},
\end{eqnarray*}
so that $\tilde{F}^N\left(\mathcal{Z}_P^N(s)\right)=\mathcal{Z}_C^N(s)$. 
As we get $(\ref{MN1_asymptotic_app})$,  $\left[N\cdot \mathcal{M}^N\right]_t$ is asymptotic to 
\begin{eqnarray*}
\int_0^t \kappa_3 \left| N\left[F^N\left(Z^N(s)\right) - \tilde{F}^N\left(Z_P^N(s)\right) \right] 
+ \frac{d \tilde{F}^N\left(\mathcal{Z}_P^N(s) \right) }{d \mathcal{Z}_P^N(s)} \mathcal{E}^N(s) 
\right| \,ds
+  \int_0^t \kappa_4 \left|\mathcal{E}^N(s)\right| \,ds. \label{MN2_asymptotic_app} 
\end{eqnarray*}

Next, we show that
\begin{eqnarray}
\int_0^t N \left[ F^N\left( Z^N(s) \right) - \tilde{F}^N \left( Z_P^N(s) \right)  \right] \,ds &\longrightarrow& 0, \label{ftil_lim_app}
\end{eqnarray}
as $N\to\infty$. Denoting
\begin{eqnarray}
A^N(z_P) &=& Z_{E_T}^N-Z_{S_T}^N+z_P+\frac{K_d}{N} \label{a_eq}\\
B^N(z_P)&=& Z_{S_T}^N-z_P, \label{b_eq}
\end{eqnarray}
we have
\begin{eqnarray}
&& N\left( F^N(z)-\tilde{F}^{N}(z_P) \right) =
 -z_S - N\left(\frac{A^N(z_P) - \sqrt{ A^N(z_P)^2+\frac{4}{N} K_d B^N(z_P) }}{2}\right) \label{estim_app}\\
&&\qquad = \left[-z_S+\frac{K_dB^N(z_P)}{A^N(z_P)} \right] \nonumber \\
&&\qquad+ \left[-\frac{K_dB^N(z_P)}{A^N(z_P)} + \frac{2K_d B^N(z_P)}{A^N(z_P)+ \sqrt{ A^N(z_P)^2+\frac{4}{N} K_d B^N(z_P) }} \right] \nonumber\\
&&\qquad = \left[-z_S+\frac{K_dB^N(z_P)}{A^N(z_P)} \right] 
+ \frac{K_dB^N(z_P)}{A^N(z_P)} \cdot \frac{-\frac{4}{N}\frac{K_dB^N(z_P)}{A^N(z_P)^2} }{\left(1+\sqrt{1+\frac{4}{N}\frac{K_dB^N(z_P)}{A^N(z_P)^2}}\right)^2}. \nonumber
\end{eqnarray}
The second term on the right is of order $\frac{1}{N}$ in $(\ref{estim_app})$. The integral of the first term in $(\ref{estim_app})$ becomes
\begin{eqnarray*}
\int_0^t \left[-Z_S^N(s)+\frac{K_dB^N\left(Z_P^N(s)\right)}{A^N\left(Z_P^N(s)\right)} \right] \,ds
&=& \int_0^t \left[ -Z_S^N(s) + \frac{\kappa_2\left(Z_{S_T}^N-Z_P^N(s)\right)}{\kappa_1\left(Z_{E_T}^N-Z_{S_T}^N+Z_P^N(s)+\frac{K_d}{N}\right)} \right]\,ds,
\end{eqnarray*}
and this converges to $0$ as $N\to\infty$ using $(\ref{slim_eq_app})$ and $(\ref{slim_app})$, which shows $(\ref{ftil_lim_app})$.

Using $\tilde{F}^N(z_P)\to F(z_P)\equiv Z_{S_T}-z_P$ and $\mathcal{Z}_P^N\to Z_P$,
\begin{eqnarray}
\frac{d \tilde{F}^N \left( \mathcal{Z}_P^N(s) \right)}{d\mathcal{Z}_P^N(s)} &\longrightarrow& \frac{d F\left(Z_P(s) \right) }{d Z_P(s)} =-1, \label{dftil_lim_app}
\end{eqnarray}
as $N\to\infty$.
Therefore, using the martingale central limit theorem,  $N\cdot \mathcal{M}^{N}\Rightarrow \mathcal{M}$ as $N\to\infty$, which is a Gaussian
process with its quadratic variation
\begin{eqnarray*}
\left[\mathcal{M}\right]_t &=& \int_0^t \left(\kappa_3+\kappa_4\right) \left|\mathcal{E}(s)\right| \,ds,
\end{eqnarray*}
where $\mathcal{E}^N(s)\Rightarrow\mathcal{E}(s)$ as $N\to\infty$. 
As we derive $(\ref{E1N_app})$, we can derive an equation for $\mathcal{E}^N(t)$ by replacing $\mathbb{E}^N$, $\mathbb{M}^N$, $\bar{F}^N$, and $\mathbb{Z}_P^N$ with $\mathcal{E}^N$, $\mathcal{M}^N$, $\tilde{F}^N$, and $\mathcal{Z}_P^N$, respectively. 
Then, $\mathcal{E}^N$ is asymptotically equal to
\begin{eqnarray}
\mathcal{E}^N(t) &\approx& 
\int_0^t \kappa_3 \left\{ N \left[ F^N\left( Z^N(s) \right) - \tilde{F}^N \left( Z_P^N(s) \right)  \right] 
+\frac{d \tilde{F}^N \left( \mathcal{Z}_P^N(s) \right)}{d\mathcal{Z}_P^N(s)} \mathcal{E}^N(s) \right\} \,ds \label{E2N_app}\\
&& - \int_0^t \kappa_4 \mathcal{E}^N(s) \,ds + N\cdot \mathcal{M}^{N}(t) . \nonumber
\end{eqnarray}
Using $(\ref{ftil_lim_app})$ and $(\ref{dftil_lim_app})$, $(\ref{E2N_app})$ converges in distribution to
\begin{eqnarray*}
\mathcal{E}(t) &=&  \int_0^t \sqrt{\left(\kappa_3+\kappa_4\right) \left|\mathcal{E}(s)\right|} \,dW(s)
- \int_0^t \left(\kappa_3+\kappa_4\right) \mathcal{E}(s) \,ds, 
\end{eqnarray*}
as $N\to\infty$ where $W$ is a standard Brownian motion. Again, we approximate $\mathcal{E}^{N_0}(t) \approx \mathcal{E}(t)$ as suggested in \cite{Kang:2014:CLT} and thus we get
\begin{eqnarray*}
X_P(t) &\approx& N_0\mathcal{Z}_P(N_0^{-3}t) + \mathcal{E}(N_0^{-3}t)\\
%&=& N_0\mathcal{Z}_P(N_0^{-3}t) .
\end{eqnarray*}
Since $\mathcal{E}(0)=0$ and diffusion and drift terms are proportional to $\mathcal{E}(s)$, $\mathcal{E}(t)=0$, which indicates that $X_P(t) = N_0\mathcal{Z}_P(N_0^{-3}t) + o(1)$. \\

\textbf{Acknowledgment} We are grateful to the MBI for supporting our attendance at the workshop in  2015, where collaboration for this work began.  We also thank Wanmo Kang for valuable discussion. This work was supported by the National Research Foundation of Korea grant N01160447 (JKK), KAIST Research Allowance grant G04150020 (JKK), the TJ Park Science Fellowship of POSCO TJ Park Foundation (JKK), National Science Foundation grant DMS-1318886 (GR), DMS-1620403 (HWK), UMBC KAN3STRT (HWK), and National Science Foundation grant DMS-0931642 to the Mathematical Biosciences Institute (JKK, GR, HWK).

\bibliographystyle{siamplain}
\bibliography{bibmulti}

\begin{thebibliography}{10}

\bibitem{Agarwal2012}
{\sc A.~Agarwal, R.~Adams, G.~C. Castellani, and H.~Z. Shouval}, {\em On the
  precision of quasi steady state assumptions in stochastic dynamics}, J. Chem.
  Phys., 137 (2012).

\bibitem{Anderson:2011:EAT}
{\sc D.~F. Anderson, A.~Ganguly, and T.~G. Kurtz}, {\em Error analysis of
  tau-leap simulation methods}, Ann. Appl. Probab., 21 (2011), pp.~2226--2262.

\bibitem{Anderson:2012:MMC}
{\sc D.~F. Anderson and D.~J. Higham}, {\em Multilevel monte carlo for
  continuous time markov chains, with applications in biochemical kinetics},
  SIAM Multiscale Model. Simul., 10 (2012), pp.~146--179.

\bibitem{Anderson:2012:WEA}
{\sc D.~F. Anderson and M.~Koyama}, {\em Weak error analysis of numerical
  methods for stochastic models of population processes}, SIAM Multiscale
  Model. Simul., 10 (2012), pp.~1493--1524.

\bibitem{Ball:2006:AAM}
{\sc K.~Ball, T.~G. Kurtz, L.~Popovic, and G.~Rempala}, {\em Asymptotic
  analysis of multiscale approximations to reaction networks}, Ann. Appl.
  Probab., 16 (2006), pp.~1925--1961.

\bibitem{Barik2008}
{\sc D.~Barik, M.~R. Paul, W.~T. Baumann, Y.~Cao, and J.~J. Tyson}, {\em
  Stochastic simulation of enzyme-catalyzed reactions with disparate
  timescales}, Biophys. J., 95 (2008), pp.~3563--3574.

\bibitem{bennett2009absolute}
{\sc B.~D. Bennett, E.~H. Kimball, M.~Gao, R.~Osterhout, S.~J. Van~Dien, and
  J.~D. Rabinowitz}, {\em Absolute metabolite concentrations and implied enzyme
  active site occupancy in {E}scherichia coli}, Nat. Chem. Biol., 5 (2009),
  pp.~593--599.

\bibitem{berglund2003geometric}
{\sc N.~Berglund and B.~Gentz}, {\em Geometric singular perturbation theory for
  stochastic differential equations}, J. Differential Equations, 191 (2003),
  pp.~1--54.

\bibitem{Bundschuh2003a}
{\sc R.~Bundschuh, F.~Hayot, and C.~Jayaprakash}, {\em The role of dimerization
  in noise reduction of simple genetic networks}, J. Theor. Biol., 220 (2003),
  pp.~261--269.

\bibitem{Cai2007}
{\sc X.~Cai and X.~Wang}, {\em {Stochastic modeling and simulation of gene
  networks-a review of the state-of-the-art research on stochastic
  simulations}}, IEEE Signal Process. Mag., 24 (2007), pp.~27--36.

\bibitem{Cao2005}
{\sc Y.~Cao, D.~T. Gillespie, and L.~R. Petzold}, {\em The slow-scale
  stochastic simulation algorithm}, J. Chem. Phys., 122 (2005).

\bibitem{coifman2008diffusion}
{\sc R.~R. Coifman, I.~G. Kevrekidis, S.~Lafon, M.~Maggioni, and B.~Nadler},
  {\em Diffusion maps, reduction coordinates, and low dimensional
  representation of stochastic systems}, SIAM Multiscale Model. Simul., 7
  (2008), pp.~842--864.

\bibitem{Cotter:2016:CAE}
{\sc S.~L. Cotter}, {\em Constrained approximation of effective generators for
  multiscale stochastic reaction networks and application to conditioned path
  sampling}, J. Comput. Phys., 323 (2016), pp.~265--282,
  \href{http://dx.doi.org/10.1016/j.jcp.2016.07.035}
  {doi:10.1016/j.jcp.2016.07.035},
  \url{http://dx.doi.org/10.1016/j.jcp.2016.07.035}.

\bibitem{Cotter:2011:CAM}
{\sc S.~L. Cotter, K.~C. Zygalakis, I.~G. Kevrekidis, and R.~Erban}, {\em A
  constrained approach to multiscale stochastic simulation of chemically
  reacting systems}, J. Chem. Phys., 135 (2011), p.~094102.

\bibitem{Durrett:2012:ESP}
{\sc R.~Durrett}, {\em Essentials of stochastic processes}, Springer Science \&
  Business Media, 2012.

\bibitem{E2005a}
{\sc W.~E, D.~Liu, and E.~Vanden-Eijnden}, {\em {Nested stochastic simulation
  algorithm for chemical kinetic systems with disparate rates.}}, J. Chem.
  Phys., 123 (2005), p.~194107.

\bibitem{erban2009stochastic}
{\sc R.~Erban and S.~J. Chapman}, {\em Stochastic modelling of
  reaction--diffusion processes: algorithms for bimolecular reactions}, Phys.
  Biol., 6 (2009), p.~046001.

\bibitem{Ethier:1986:MPC}
{\sc S.~N. Ethier and T.~G. Kurtz}, {\em Markov processes: characterization and
  convergence}, vol.~282, John Wiley \& Wiley, 1986.

\bibitem{Ganguly:2015:EBS}
{\sc A.~Ganguly, D.~Altintan, and H.~Koeppl}, {\em Error bound and simulation
  algorithm for piecewise deterministic approximations of stochastic reaction
  systems}, in American Control Conference (ACC), 2015, IEEE, 2015,
  pp.~787--792.

\bibitem{Ganguly:2015:JAS}
{\sc A.~Ganguly, D.~Altintan, and H.~Koeppl}, {\em Jump-diffusion approximation
  of stochastic reaction dynamics: error bounds and algorithms}, SIAM
  Multiscale Model. Simul., 13 (2015), pp.~1390--1419.

\bibitem{Gillespie2007}
{\sc D.~T. Gillespie}, {\em Stochastic simulation of chemical kinetics}, Ann.
  Rev. Phys. Chem., 58 (2007), pp.~35--55.

\bibitem{gillespie2009JPC}
{\sc D.~T. Gillespie}, {\em Deterministic limit of stochastic chemical
  kinetics}, J. Phys. Chem. B, 113 (2009), pp.~1640--1644.

\bibitem{givon2007strong}
{\sc D.~Givon}, {\em Strong convergence rate for two-time-scale jump-diffusion
  stochastic differential systems}, SIAM Multiscale Model. Simul., 6 (2007),
  pp.~577--594.

\bibitem{goeke2014constructive}
{\sc A.~Goeke and S.~Walcher}, {\em A constructive approach to quasi-steady
  state reductions}, J. Math. Chem., 52 (2014), pp.~2596--2626.

\bibitem{Goutsias2005}
{\sc J.~Goutsias}, {\em Quasiequilibrium approximation of fast reaction
  kinetics in stochastic biochemical systems}, J. Chem. Phys., 122 (2005).

\bibitem{Gupta:2013:UEP}
{\sc A.~Gupta and M.~Khammash}, {\em Unbiased estimation of parameter
  sensitivities for stochastic chemical reaction networks}, SIAM J. Sci.
  Comput., 35 (2013), pp.~A2598--A2620.

\bibitem{Gupta:2014:SAS}
{\sc A.~Gupta and M.~Khammash}, {\em Sensitivity analysis for stochastic
  chemical reaction networks with multiple time-scales}, Electron. J. Probab.,
  19 (2014), pp.~1--53.

\bibitem{haseltine2005origins}
{\sc E.~L. Haseltine and J.~B. Rawlings}, {\em On the origins of approximations
  for stochastic chemical kinetics}, J. Chem. Phys., 123 (2005), p.~164115.

\bibitem{Hepp:2015:AHS}
{\sc B.~Hepp, A.~Gupta, and M.~Khammash}, {\em Adaptive hybrid simulations for
  multiscale stochastic reaction networks}, J. Chem. Phys., 142 (2015),
  p.~034118.

\bibitem{Hu:2014:SAR}
{\sc J.~Hu, H.-W. Kang, and H.~G. Othmer}, {\em Stochastic analysis of
  reaction--diffusion processes}, Bull. Math. Biol., 76 (2014), pp.~854--894.

\bibitem{isaacson2006incorporating}
{\sc S.~A. Isaacson and C.~S. Peskin}, {\em Incorporating diffusion in complex
  geometries into stochastic chemical kinetics simulations}, SIAM J. Sci.
  Comput., 28 (2006), pp.~47--74.

\bibitem{jahnke2011reduced}
{\sc T.~Jahnke}, {\em On reduced models for the chemical master equation}, SIAM
  Multiscale Model. Simul., 9 (2011), pp.~1646--1676.

\bibitem{Kang:2012:MAH}
{\sc H.-W. Kang}, {\em A multiscale approximation in a heat shock response
  model of {E}. coli}, BMC Syst. Biol., 6 (2012), p.~143.

\bibitem{Kang:2013:STM}
{\sc H.-W. Kang and T.~G. Kurtz}, {\em Separation of time-scales and model
  reduction for stochastic reaction networks}, Ann. Appl. Probab., 23 (2013),
  pp.~529--583.

\bibitem{Kang:2014:CLT}
{\sc H.-W. Kang, T.~G. Kurtz, and L.~Popovic}, {\em Central limit theorems and
  diffusion approximations for multiscale {M}arkov chain models}, Ann. Appl.
  Probab., 24 (2014), pp.~721--759.

\bibitem{Kang:2012:NMC}
{\sc H.-W. Kang, L.~Zheng, and H.~G. Othmer}, {\em A new method for choosing
  the computational cell in stochastic reaction--diffusion systems}, J. Math.
  Biol., 65 (2012), pp.~1017--1099.

\bibitem{Kepler2001}
{\sc T.~B. Kepler and T.~C. Elston}, {\em Stochasticity in transcriptional
  regulation: Origins, consequences, and mathematical representations},
  Biophys. J., 81 (2001), pp.~3116--3136.

\bibitem{kim2016IET}
{\sc J.~K. Kim}, {\em Protein sequestration versus hill-type repression in
  circadian clock models}, IET Syst. Biol., 10 (2016), pp.~125--135(10).

\bibitem{Kim2012}
{\sc J.~K. Kim and D.~B. Forger}, {\em A mechanism for robust circadian
  timekeeping via stoichiometric balance}, Mol. Syst. Biol., 8 (2012).

\bibitem{kim2014validity}
{\sc J.~K. Kim, K.~Josi{\'c}, and M.~R. Bennett}, {\em The validity of
  quasi-steady-state approximations in discrete stochastic simulations},
  Biophys. J., 107 (2014), pp.~783--793.

\bibitem{kim2015relationship}
{\sc J.~K. Kim, K.~Josi{\'c}, and M.~R. Bennett}, {\em The relationship between
  stochastic and deterministic quasi-steady state approximations}, BMC Syst.
  Biol., 9 (2015), p.~87.

\bibitem{Kim2014}
{\sc J.~K. Kim, Z.~P. Kilpatrick, M.~R. Bennett, and K.~Josi\'{c}}, {\em
  {Molecular mechanisms that regulate the coupled period of the mammalian
  circadian clock}}, Biophys. J., 106 (2014), pp.~2071--2081.

\bibitem{Kurtz:1972:RSD}
{\sc T.~G. Kurtz}, {\em The relationship between stochastic and deterministic
  models for chemical reactions}, J. Chem. Phys., 57 (1972), pp.~2976--2978.

\bibitem{Kurtz:1978:SAT}
{\sc T.~G. Kurtz}, {\em Strong approximation theorems for density dependent
  markov chains}, Stoch. Proc. Appl., 6 (1978), pp.~223--240.

\bibitem{Kurtz:1981:APP}
{\sc T.~G. Kurtz}, {\em Approximation of population processes}, vol.~36, SIAM,
  1981.

\bibitem{Kurtz:1992:AMP}
{\sc T.~G. Kurtz}, {\em Averaging for martingale problems and stochastic
  approximation}, in Applied Stochastic Analysis, vol.~177, Springer, 1992,
  pp.~186--209.

\bibitem{liu2010analysis}
{\sc D.~Liu}, {\em Analysis of multiscale methods for stochastic dynamical
  systems with multiple time scales}, SIAM Multiscale Model. Simul., 8 (2010),
  pp.~944--964.

\bibitem{lotstedt2006dimensional}
{\sc P.~L{\"o}tstedt and L.~Ferm}, {\em Dimensional reduction of the
  {F}okker-{P}lanck equation for stochastic chemical reactions}, SIAM
  Multiscale Model. Simul., 5 (2006), pp.~593--614.

\bibitem{Macnamara2008}
{\sc S.~MacNamara, A.~M. Bersani, K.~Burrage, and R.~B. Sidje}, {\em Stochastic
  chemical kinetics and the total quasi-steady-state assumption: Application to
  the stochastic simulation algorithm and chemical master equation}, J. Chem.
  Phys., 129 (2008).

\bibitem{michelotti2013binning}
{\sc M.~D. Michelotti, M.~T. Heath, and M.~West}, {\em Binning for efficient
  stochastic multiscale particle simulations}, SIAM Multiscale Model. Simul.,
  11 (2013), pp.~1071--1096.

\bibitem{pelevs2006reduction}
{\sc S.~Pele{\v{s}}, B.~Munsky, and M.~Khammash}, {\em Reduction and solution
  of the chemical master equation using time scale separation and finite state
  projection}, J. Chem. Phys., 125 (2006), p.~204104.

\bibitem{Pfaffelhuber:2015:SLS}
{\sc P.~Pfaffelhuber and L.~Popovic}, {\em Scaling limits of spatial
  compartment models for chemical reaction networks}, Ann. Appl. Probab., 25
  (2015), pp.~3162--3208.

\bibitem{Rao2003}
{\sc C.~V. Rao and A.~P. Arkin}, {\em Stochastic chemical kinetics and the
  quasi-steady-state assumption: Application to the {G}illespie algorithm}, J.
  Chem. Phys., 118 (2003), pp.~4999--5010.

\bibitem{salis2005equation}
{\sc H.~Salis and Y.~N. Kaznessis}, {\em An equation-free probabilistic
  steady-state approximation: Dynamic application to the stochastic simulation
  of biochemical reaction networks}, J. Chem. Phys., 123 (2005), p.~214106.

\bibitem{Thomas2011}
{\sc P.~Thomas, A.~V. Straube, and R.~Grima}, {\em {Communication: limitations
  of the stochastic quasi-steady-state approximation in open biochemical
  reaction networks.}}, J. Chem. Phys., 135 (2011), p.~181103,
  \href{http://dx.doi.org/10.1063/1.3661156} {doi:10.1063/1.3661156}.

\bibitem{Thomas2012}
{\sc P.~Thomas, A.~V. Straube, and R.~Grima}, {\em The slow-scale linear noise
  approximation: an accurate, reduced stochastic description of biochemical
  networks under timescale separation conditions}, BMC Syst. Biol., 6 (2012).

\bibitem{tikhonov1952systems}
{\sc A.~N. Tikhonov}, {\em Systems of differential equations containing small
  parameters in the derivatives}, Mat. Sb. (N.S.), 31(73) (1952), pp.~575--586.

\bibitem{van1985elimination}
{\sc N.~G. Van~Kampen}, {\em Elimination of fast variables}, Phys. Rep., 124
  (1985), pp.~69--160.

\bibitem{vanden2003fast}
{\sc E.~Vanden-Eijnden}, {\em Fast communications: Numerical techniques for
  multi-scale dynamical systems with stochastic effects}, Commun. Math. Sci., 1
  (2003), pp.~385--391.

\end{thebibliography}

\end{document}